\documentclass{statsoc}

\usepackage[a4paper]{geometry}
\usepackage{graphicx,graphics,bm,epstopdf}
\usepackage[textwidth=8em,textsize=small]{todonotes}
\usepackage{amsmath,amsfonts,amssymb,hyperref,verbatim}
\usepackage{natbib}
\usepackage[para,online,flushleft]{threeparttable}
\title[DW regression]{Discrete Weibull generalised additive model: an application to count fertility data}
\author[]{Alina Peluso}
\address{Department of Mathematics, Brunel University London, Uxbridge UB8 3PH, UK.}
\author[]{Veronica Vinciotti}
\address{Department of Mathematics, Brunel University London, Uxbridge UB8 3PH, UK.}
\email{veronica.vinciotti@brunel.ac.uk}
\author[]{Keming Yu}
\address{Department of Mathematics, Brunel University London, Uxbridge UB8 3PH, UK.}

\begin{document}

\begin{abstract}
Fertility plans, measured by the number of planned children, have been found to be affected by education and family background via complex tail dependencies. This challenge was previously met with the use of non-parametric jittering approaches. This paper shows how a novel generalized additive model based on a discrete Weibull distribution provides partial effects of the covariates on fertility plans which are comparable to jittering, without the inherent drawback of crossing conditional quantiles. The model has some additional desirable features: both over- and under-dispersed data can be modelled by this distribution, the conditional quantiles have a simple analytic form and the likelihood is the same of that of a continuous Weibull distribution with interval-censored data. The latter means that efficient implementations are already available, in the R package {\tt gamlss}, for a range of models and inferential procedures, and at a fraction of the time compared to the jittering and COM-Poisson approaches, showing potential for the wide applicability of this approach to the modelling of count data.
\end{abstract}

\section{Introduction}

Fertility plans measured by the number of planned children, or ideal fertility, have been previously found to be affected by education and family background \citep{knodel1973desired,pritchett1994desired}. In a recent study in Mexico, \cite{miranda08} showed how the dependency of ideal fertility on education and family background is complex, with effects mostly at the tail of the distribution, and how ideal fertility is typically under-dispersed relative to Poisson. In this paper, using the latest data from the Mexican National Survey of Demographic Dynamics (ENADID from its acronym in Spanish, \cite{inegi14}), we propose a novel regression model to discover determinants of planned fertility and quantify this dependency.

Methods to address questions such as this fall in the general area of regression analysis of count data, with many applications ranging from healthcare, biology, social science, marketing and crime data analyses (\cite{cameron2013regression,hilbe2014modeling}). Amongst these methods, generalised linear models \citep{nelder1972generalized} are popular in the parametric literature. Here, the conditional distribution of the response variable given the predictors is assumed to follow a specified distribution, with the conditional mean linked to the predictors via a regression model. For example, Poisson regression assumes that the conditional distribution is Poisson with a conditional mean regressed on the covariates through the log link function. Although Poisson regression is fundamental to the regression analysis of count data, it is often of limited use for real data, due to its property of equal mean and variance. Real data usually presents over-dispersion relative to Poisson or the opposite case of under-dispersion. Negative Binomial regression is widely considered as the default choice for data that are over-dispersed relative to Poisson, although other options, such as the Poisson-inverse Gaussian model \citep{willmot87}, are available. However, Negative Binomial regression as well as the Poisson-inverse Gaussian model, cannot deal with data that are under-dispersed relative to Poisson. These can arise in various applications, such as in cases where the data are pre-processed due to confidentiality issues \citep{kadane06}. There have been some attempts to extend Poisson-based models to include also under dispersion, such as the generalised Poisson regression model \citep{consul92}, Conway-MaxwellPoisson (COM-Poisson) regression \citep{sellers2008flexible}, extended Poisson processes models \citep{smith16} or hyper-Poisson regression models \citep{saez2013hyper}. These models are all modifications of a Poisson model and have been shown to be rather complex and computationally intensive in practice \citep{chanialidis17}.

At the other spectrum of parametric approaches, quantile regression approaches focus on modelling individual quantiles of the distribution and linking these to the predictors via a regression model. Of particular notice for discrete responses are the quantile regression models for binary and multinomial response of \cite{manski85} and \cite{horowitz92}, and the median regression approach with ordered response of \cite{lee92}. For a general discrete response, the literature on quantile regression for counts is mainly dominated by the jittering approach of \cite{machado05}, which was also rephrased in a Bayesian framework by \cite{lee10} in the context of an environmental epidemiology study. In these approaches, the fitted regression parameters are specific to the selected conditional quantile, thanks to the use of quantile-specific loss functions. Performing inference across a range of quantiles provides a global picture of the conditional distribution of the response variable, without having to specify the parametric form of the conditional distribution. This has proven to be rather useful in practice, particularly in cases where the relationship between response and predictors is complex. This was in fact found in the planned fertility dataset of \cite{miranda08}, whereby a jittering approach revealed effects mostly at the tails of the conditional distribution.

Quantile regression approaches, however, suffer from some drawbacks: inference has to be made for each individual quantile, separate quantiles may cross and, in the case of jittering, the underlying uniform random sampling can generate instability in the estimation. The parametric literature, on the other hand, has addressed more complex dependencies by developing new distributions with additional parameters, e.g. the generalised Gamma approach of \cite{noufaily2013parametric} for continuous responses, and/or by adopting more flexible non-linear regression models that can link all parameters of the distribution to the covariates, most notably the generalized additive models for location, scale and shape (GAMLSS) of \cite{rigby05}. This paper fits within this literature. In particular, we introduce a generalised additive discrete Weibull regression model. The discrete Weibull distribution itself was originally developed by \cite{nakagawa1975discrete} as a discretized form of the continuous Weibull distribution, which is popular in the survival analysis and failure time studies. Since then, aside from some early work on parameter estimation \citep{khan1989estimating,kulasekera1994approximate}, and some limited use in applied contexts \citep{englehardt2011discrete,englehardt2012methods}, there are not many other contributions in the literature. Recently, we have introduced this distribution in a simple linear regression context  \citep{kalktawi2015simple,haselimashhadi17}, showing a number of desirable features: it can model both over- and under-dispersed data, without being restricted to either of the two, and the conditional quantiles have a simple analytical form.  Moreover, since the likelihood from a discrete Weibull model is the same as that of a continuous Weibull distribution with interval-censored data, efficient implementations of more complex models, such as non-linear models, mixed models and mixture models, are already available in the R package {\tt gamlss} \citep{rigby05}.

We present the discrete Weibull distribution in Section \ref{sec:distribution} and the novel generalised additive model in Section \ref{sec:model}. In Section \ref{sec:simulation}, we assess the performance of the proposed model on a simulation study, against that of the jittering approach and existing parametric approaches. Finally, in Section \ref{sec:realdata}, we show how the discrete Weibull generalized additive model selected on the real data returns partial effects of planned fertility similar to those of the jittering approach, without the inherent drawback of quantile crossing, and is comparable to the fitting of COM-Poisson, at a fraction of the time.

\section{Discrete Weibull} \label{sec:distribution}
In this section, we report some important results on the discrete Weibull distribution which will be used later on in the paper.
\subsection{The distribution}
If $ Y $ follows a (type 1) DW distribution \citep{nakagawa1975discrete}, then the cumulative distribution function of $ Y $ is given by
	
\begin{equation*}\label{eq:cdf discrete weibull}
F(y;q,\beta)=
\begin{cases}
1-q^{(y+1)^{\beta}} & \mbox{\text{for} $y=0,1,2,3, \ldots$} \\
0 & \mbox{otherwise}
\end{cases}
\end{equation*}
and its probability mass function by
\begin{equation}\label{eq:pmf discrete weibull}
f(y;q,\beta)=	
\begin{cases}
q^{y^{\beta}}-	q^{(y+1)^{\beta}} & \mbox{\text{for} $y=0,1,2,3, \ldots$} \\
0 & \mbox{otherwise}
\end{cases}
\end{equation}
with the parameters $ 0<q<1 $ and $ \beta>0 $. Since $f(0)=1-q$, the parameter $q$ is directly related to the percentage of zeros.


\subsection{Moments and quantiles}

It can be shown that for a DW distribution:
\begin{eqnarray}
\label{eq:expectedDW}
E(Y)&=&\sum_{y=1}^{\infty} q^{y^{\beta}}\\
E(Y^2)&=&\sum_{y=1}^{\infty} (2y-1)q^{y^{\beta}} = 2\sum_{y=1}^{\infty} yq^{y^{\beta}}-E(Y), \nonumber
\end{eqnarray}
for which there are no closed form expressions, but numerical approximations can be obtained on a truncated support \citep{barbiero2013package}.

As for quantiles, the $\tau$ quantile of a DW distribution is given by the smallest integer $\mu_{(\tau)}$ for which $P(Y \le \mu_{(\tau)})= 1-q^{(y+1)^{\beta}} \ge \tau$. This gives
\begin{equation}
\mu_{(\tau)}= \Big\lceil{\mu}_{(*\tau)}\Big\rceil=\Big\lceil \biggl(\dfrac{\log(1-\tau)}{\log(q)} \biggr)^{1/{\beta}}-1 \Big\rceil,
\label{eq:quantileDW}
\end{equation}
with $\lceil \cdot \rceil$ the ceiling function. From this
\begin{equation}
\log(\mu_{(*\tau)}+1)=\dfrac{1}{\beta}\log(\!-\!\log(1-\tau))-\dfrac{1}{\beta}\log(\!-\!\log(q)).
\label{eq:logquantileDW}
\end{equation}
Given that $Y$ is non negative, the quantile is defined only for $\tau \ge 1-q$. As a special case, the median of a DW distribution is given by:
\begin{equation}
\mu_{(0.5)}=\Big\lceil\biggl(-\displaystyle\frac{\ln(2)}{\ln(q)}\biggr)^{\frac{1}{\beta}}-1 \Big\rceil.
\label{eq:medianDW}
\end{equation}
Thus the quantiles of a DW distribution are given by simple, analytical formulae.

\subsection{Likelihood and link with continuous Weibull}

There is a natural link between the DW distribution and the continuous Weibull distribution with interval censored data. The DW distribution was in fact developed as a discretized form of the continuous Weibull distribution \citep{chakraborty2015generating}. In particular, let $Y$ be a random variable distributed as a continuous Weibull, with probability density function and cumulative density function :
\begin{align}
\begin{split}
\nonumber
 f_W(y; q,\beta) &= \beta \, \log(q) \, y^{(\beta-1)} \, \exp(-y^{\beta} \, \log(q)) \qquad  y\geq0 \\ \nonumber
 F_W(y; q,\beta) &= 1 - \exp(-y^{\beta} \, \log(q)),
\end{split}
\end{align}
respectively.
Then one can show that
\begin{align}
\begin{split}
 f(y)=F_W(y+1)-F_W(y) \quad y=0,1,\ldots \nonumber
\end{split}
\end{align}
where $f(y)$ is the DW probability mass function of equation (\ref{eq:pmf discrete weibull}). From this
\begin{align}
\begin{split}
\int^{y+1}_{y}f_{\textrm{W}}(t)dt&=f(y).  \nonumber
\end{split}
\end{align}
Thus the likelihood of a continuous Weibull distribution with interval censored data is equal to that of a DW distribution, i.e.
\begin{align}
\begin{split}
\prod_{i=1}^n f(y_i)=\prod_{i=1}^n \left(F_W(y_i+1)- F_W(y_i) \right).  \nonumber
\end{split}
\end{align}

\subsection{DW accounts for over and under dispersion}

Dispersion in count data is formally defined in relation to a specified model being fitted to the data \citep{cameron2013regression}. In particular, let
\begin{equation*} \label{eq:varratio}
VR=\dfrac{\text{observed variance}}{\text{theoretical variance}}.
\end{equation*}
So VR is the ratio between the observed  variance from the data and the theoretical variance from the model. Then the data are said to be over-/equi-/under- dispersed relative to the fitted model if the observed variance is larger/equal/smaller than the theoretical variance specified by the model, respectively.
It is common to refer to dispersion relative to Poisson. In that case, the variance of the model is estimated by the sample mean. Thus, over-/equi-/under- dispersion relative to Poisson refers to cases where the sample variance is larger/equal/smaller than the sample mean, respectively. Since the theoretical variance of a NB is always greater than its mean, the NB regression model is the natural choice for data that are over-dispersed relative to Poisson. However, crucially, NB cannot handle under-dispersed data.

\begin{figure}
\centering
\includegraphics[scale=0.3]{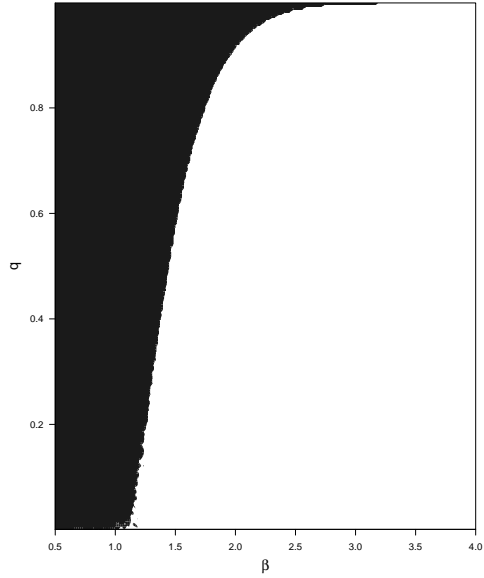}
\caption{\label{fig:observed_and_fitted_variances} Ratio of observed and theoretical variance from a Poisson model, calculated from data simulated by a DW$(q,\beta)$.}
\end{figure}

In contrast to this, \cite{kalktawi2015simple} show how a DW distribution can handle data that are both over- and under- dispersed relative to Poisson. In particular, Figure \ref{fig:observed_and_fitted_variances} shows how the DW can capture both cases of under-, equi- and over- dispersion relative to Poisson. Specifically, the white area corresponds to values of dispersion less than 1, i.e. under-dispersed relative to Poisson, whereas the black area corresponds to over-dispersion.
Moreover, the plot shows that:
\begin{itemize}
\item $0<\beta \leq 1$ is a case of over-dispersion, regardless of the value of $q$.
\item $\beta \geq 3$ is a case of under-dispersion, regardless of the value of $q$. In fact, the DW distribution approaches the Bernoulli distribution with mean $p$ and variance $p(1-p)$ for $\beta \rightarrow \infty$.
\item $1<\beta<3 $ leads to both cases of over and under-dispersion depending on the value of $q$.
\end{itemize}

\section{Discrete Weibull generalized additive model} \label{sec:model}
\subsection{GAMLSS formulation}
In order to capture complex dependencies between the response and the covariates, such as those that we expect in our real application on planned fertility, we propose generalized additive models to link both parameters of the distribution to the covariates.  Specifically, we assume that the response $Y$ has a conditional DW distribution, with the  parameters $q$ and $\beta$ linked to the covariates $x$ as follows
\begin{align}
\begin{split}
\log\left(\!-\!\log \left(q(x) \right) \right)&= \sum_{p=1}^P \sum_{d=0}^{D_p} \theta_{0pd} {x_p}^d + \sum_{p=1}^P\sum_{k=1}^{K_p} \theta_{pk}({x_p}-g_{pk})^{D_p} I({x_p}>g_{pk}), \\
\log\left(\beta(x) \right) &= \sum_{p=1}^P \sum_{d=0}^{D_p'} \vartheta_{0pd} {x_p}^d +
\sum_{p=1}^P \sum_{k=1}^{K_p'} \vartheta_{pk}({x_p}-g_{pk})^{D_p'} I({x_p}>g_{pk}),
\label{eq:GAMmodelDW}
\end{split}
\end{align}
where $x=(1,x_1,\ldots,x_P)$ is the vector of covariates, $D$ and $D'$ denote the polynomial degrees, $K$ and $K'$ are the number of knots for each covariate $X_p$, with $g_{pk}$ the corresponding knots, $I(\cdot)$ is the indicator function and ($\bm{\theta},\bm{\vartheta}$) is the vector of parameters to be estimated. The $\log\!-\!\log$ link in $q$ is motivated by the analytical formula for the quantile (\autoref{eq:logquantileDW}), which facilitates the interpretation of the parameters as discussed in the next subsection. Other link functions are possible, such as the logit link on $q$, as explored in \citep{haselimashhadi17} for the simple regression case.

The general formulation presented in \autoref{eq:GAMmodelDW} includes models of varying complexity, such as linear models, orthogonal polynomial basis \citep{szeg1939orthogonal} and B-splines models \citep{de1972calculating}. Rather than defining the number of knots and degrees, it is also possible to formulate the problem as a penalized regression spline \citep{wood2006generalized}. Similarly, it is possible to add random effects to each of the two regressions. Thanks to the link with the continuous Weibull likelihood described in Section \ref{sec:distribution}, inference for these models is available in the R package {\tt gamlss} under the {\tt WEIic} family \citep{stasinopoulos2007generalized}.

\begin{figure}
\centering
 \begin{minipage}[b]{0.49\textwidth}
   \includegraphics[width=\textwidth]{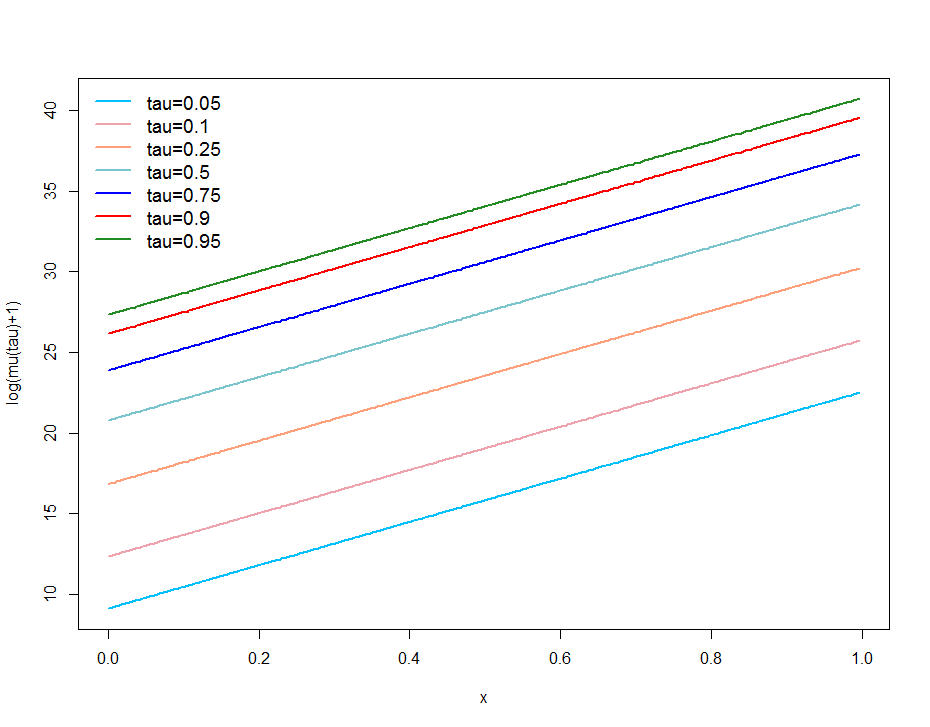}\ \centering Linear model for $q$, $\beta$ constant.
  \end{minipage}
 \hfill
  \begin{minipage}[b]{0.49\textwidth}
    \includegraphics[width=\textwidth]{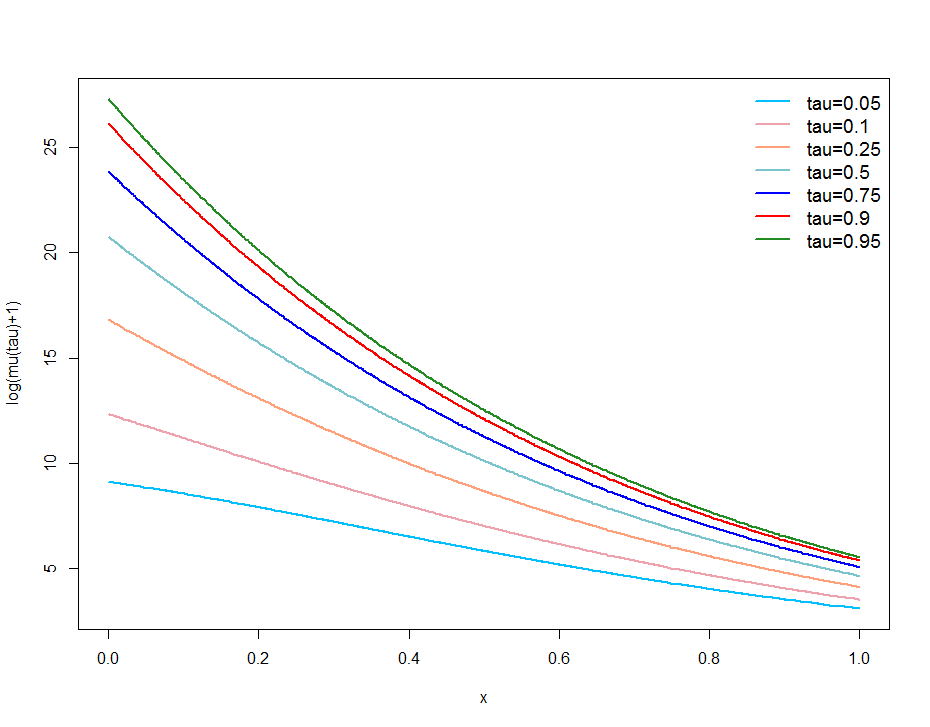}\ \centering Linear model for $q$ and $\beta$.
  \end{minipage}
 \hfill
  \begin{minipage}[b]{0.49\textwidth}
    \includegraphics[width=\textwidth]{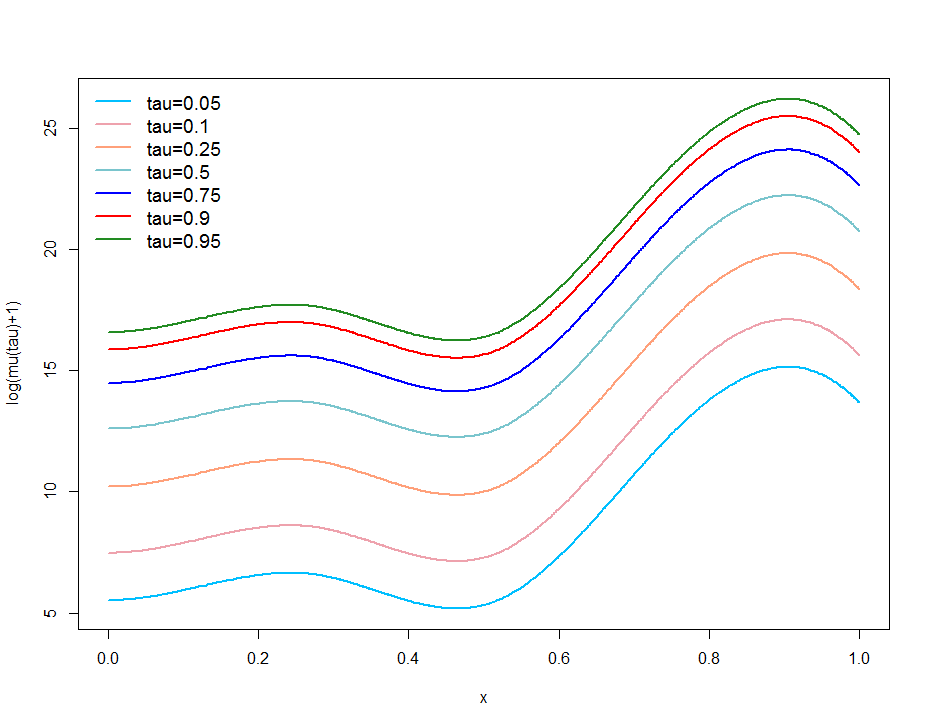}\ \centering B-spline model for $q$, $\beta$ constant.
  \end{minipage}
 \hfill
  \begin{minipage}[b]{0.49\textwidth}
    \includegraphics[width=\textwidth]{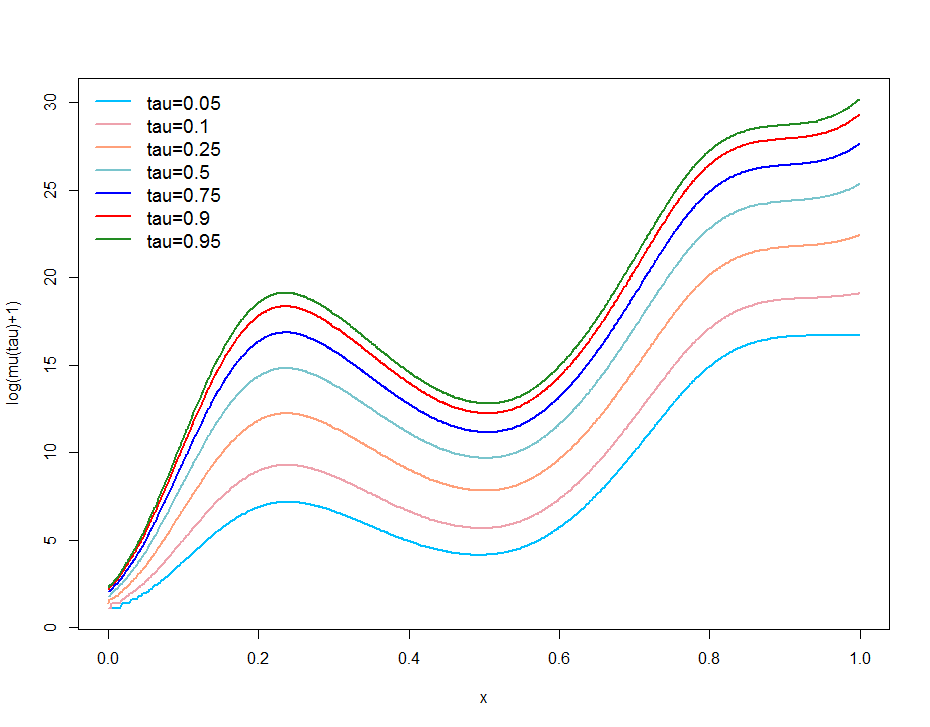}\ \centering B-spline model for $q$ and $\beta$.
  \end{minipage}
  \caption{Plot of the conditional quantiles for DW models under linear (top) and non-linear (bottom) models, and $\beta$ fixed (left) and not (right).}
\label{fig:qplotDW}
\end{figure}

Adding a link to both parameters means that conditional quantiles of various shapes and complexity can be captured. Considering one covariate $x$ only, and dropping the indices $p$ of the model for simplicity, we look closely at four cases to inspect the level of flexibility of a DW model in approximating conditional distributions.
\begin{enumerate}
\item \textbf{DW linear regression model on q(x) with $\beta$ constant.}\\
This model is specified as in \autoref{eq:GAMmodelDW} with $D=1$, $D'=0$ and no knots, i.e.:
\begin{eqnarray*}
\log\left(-\log \left(q(x) \right) \right) &=& \theta_{00}+\theta_{01}x \\
\log\left(\beta\right) &=& \vartheta_{00}.
\label{eq:GAMmodelDW_case1}
\end{eqnarray*}
The top-left plot in \autoref{fig:qplotDW} shows the case $\theta_{00}= -5 $, $\theta_{01} =-3$, $\vartheta_{00}=-1.5$. The figure plots $\log({\mu}_{(*\tau)}+1)$ from \autoref{eq:logquantileDW}. As expected by that equation, a linear model with $\beta$ constant returns log-quantiles which are linear and parallel.
\item  \textbf{DW linear regression model on q(x) and $\beta(x)$.}\\
This model is specified as in \autoref{eq:GAMmodelDW} with $D=D'=1$ and no knots, for example:
\begin{eqnarray*}
\log\left(-\log \left(q(x) \right) \right) &=& \theta_{00}+\theta_{01}x \\
\log\left(\beta(x) \right) &=& \vartheta_{00}+\vartheta_{01}x,
\label{eq:GAMmodelDW_case2}
\end{eqnarray*}
for the case of a linear model on both $q(x)$ and $\beta(x)$.
The top-right plot in \autoref{fig:qplotDW} shows the case $\theta_{00}=-5 $, $\theta_{01}=-3$, $\vartheta_{00}=-1.5$, $\vartheta_{01}=2$. This plot shows how a non-constant $\beta$ allows to obtain log-quantiles that are not parallel.
\item  \textbf{DW non-linear model for $q(x)$ with $\beta$ constant} \\
Setting $D=K=3$, $D'=K'=0$ in \autoref{eq:GAMmodelDW} leads to a B-spline model for $q(x)$ with three interior knots:
\begin{eqnarray*}
\log\left(-\log \left(q(x) \right) \right) &=& \theta_{00} + \theta_{01} x + \theta_{02} x^2 + \theta_{03} x^3 + \theta_{1}(x-g_{1})^3 I(x>g_{1}) +\\
&+& \theta_{2}(x-g_{2})^3 I(x>g_{2}) + \theta_{3}(x-g_{3})^3 I(x>g_{3}) \\
\log\left(\beta \right) &=& \vartheta_{00}.
\label{eq:GAMmodelDW_case3}
\end{eqnarray*}
Cubic splines are typically complex enough for most real applications \citep{dierckx1995curve}.
The bottom-left plot in \autoref{fig:qplotDW} shows the quantiles for the cubic spline model with $\theta_{00}=-5$, $\theta_{01}=-5$, $\theta_{02}=-6$, $\theta_{03}=-4$, $\theta_{1}=-8$, $\theta_{2}=-9$, $\theta_{3}=-8$, and $\vartheta_{00}=-1$. The cubic spline, together with the assumption of a constant $\beta$, leads to parallel and non-linear log-quantiles, as expected by \autoref{eq:logquantileDW}.
\item  \textbf{DW non-linear model for $q(x)$ and $\beta(x)$} \\
Setting $D=K=D'=K'=3$ in \autoref{eq:GAMmodelDW} leads to a B-spline model for $q(x)$ and $\beta(x)$ with three interior knots:
\begin{eqnarray*}
\log\left(-\log \left(q(x) \right) \right) &=& \theta_{00} + \theta_{01} x + \theta_{02} x^2 + \theta_{03} x^3 + \theta_{1}(x-g_{1})^3 I(x>g_{1}) +\\
&+& \theta_{2}(x-g_{2})^3 I(x>g_{2}) + \theta_{3}(x-g_{3})^3 I(x>g_{3}) \\ \nonumber
\log\left(\beta(x) \right) &=& \vartheta_{00} + \theta_{01} x + \vartheta_{02} x^2 + \vartheta_{03} x^3 + \vartheta_{1}(x-g_{1})^3 I(x>g_{1}) +\\
&+& \vartheta_{2}(x-g_{2})^3 I(x>g_{2}) + \vartheta_{3}(x-g_{3})^3 I(x>g_{3}).
\label{eq:GAMmodelDW_case4}
\end{eqnarray*}
The bottom-right plot in \autoref{fig:qplotDW} shows the quantiles for the cubic spline model with
$\theta_{00}=-5$, $\theta_{01}=-5$, $\theta_{02}=-6$, $\theta_{03}=-4$, $\theta_{1}=-8$, $\theta_{2}=-9$, $\theta_{3}=-8$, and $\vartheta_{00}=1$, $\vartheta_{01}=-1.1$, $\vartheta_{02}=-1.2$, $\vartheta_{03}=-0.5$, $\vartheta_{1}=-1.3$, $\vartheta_{2}=-1$, $\vartheta_{3}=-1.2$. The cubic spline on both parameters leads to non-parallel and non-linear log-quantiles.
\end{enumerate}

\subsection{Interpretation of DW regression coefficients and output}

After a DW regression model has been estimated, the following can be obtained:
\begin{itemize}
	\item The fitted values for the central trend of the conditional distribution, namely:
	\begin{itemize}
		\item mean: \autoref{eq:expectedDW}, as mentioned earlier, can be calculated numerically using the approximated moments of the DW \citep{barbiero2013package}.
		\item median: \autoref{eq:medianDW} can be applied. Due to the skewness, which is common for count data, the median is more appropriate than the mean.
    \end{itemize}
    \item The conditional quantile for any $ \tau $ using \autoref{eq:quantileDW}, i.e.
\[\mu_{(\tau)}(x)= \Big\lceil{\mu}_{(*\tau)}(x)\Big\rceil=\Big\lceil \biggl(\dfrac{\log(1-\tau)}{\log(q(x))} \biggr)^{1/{\beta(x)}}-1 \Big\rceil.\]
    \end{itemize}
The analytical expression of the quantiles, combined with the chosen link function, offers a way of interpreting the parameters. Considering a simple regression model on $q$ (case (a) above, \cite{kalktawi2015simple}), \autoref{eq:logquantileDW} leads to
\begin{equation*}
\log\left(\mu_{(*0.5)}(x)+1\right) =\frac{1}{\beta} \log\big(\log(2)\big)-\frac{1}{\beta} x'\bm{\theta}.
\end{equation*}
Thus, the regression parameters $\bm{\theta}$ can be interpreted in relation to the log of the median, in analogy with Poisson and NB models where the parameters are linked to the log of the mean. In particular, $\dfrac{\log\big(\log(2) \big) - \theta_0}{\beta} $ is related to the conditional median when all covariates are set to zero, whereas $\dfrac{-\theta_p}{\beta} $, $ p=1, \ldots,P $, can be related to the change in the median of the response corresponding to a one unit change of $X_p$, keeping all other covariates constant.

For more complex models, partial effects can be computed for each covariate and for each quantile as in \cite{machado05}. In particular, let $x^0$ denote the vector of predictors, where each predictor is set to their sample mean $\bar{x}$ if continuous and to 0 if dummy. Then, the effect for the regressor $x_p$ on the $\tau$ quantile of the response is calculated as the difference $\mu_{(*\tau)}(x_p^1)-\mu_{(*\tau)}(x^0)$, where $x_p^1$ is equal to $x^0$ for all entries with the exception of the $p^{th}$ entry which is increased by one unit.
	
\subsection{Model selection and diagnostic checks}
Model selection, in terms of polynomial degree and the number of interior knots, is carried out based on known model selection criteria. In this paper we will use the Akaike Information Criterion (AIC).
After fitting a DW regression model, goodness of fit is checked based on the randomized quantile residuals, as developed by \cite{dunn1996randomized} and advised in the case of non-Gaussian responses \citep{rigby05}. In particular, let
	\begin{equation*}
		\label{eq:randomizesquantileresiduals}
		{\hat r_i}=\Phi^{-1}(u_i), \qquad i=1, \ldots, n
	\end{equation*}
	where $ \Phi(.) $ is the standard normal distribution function and $ u_i $ is a uniform random variable on the interval
	\begin{equation*}
		\begin{split}
			\left( a_i , b_i \right] & = \bigg( \underset{y \uparrow y_i}{\mathrm{lim}}
			F(y;\hat{q_i},\hat{\beta_i}) , F(y_i;\hat{q_i},\hat{\beta_i}) \bigg] \\
			& \approx \bigg[ F(y_i -1;\hat{q_i},\hat{\beta_i}) ,  F(y_i;\hat{q_i},\hat{\beta_i})\bigg].
		\end{split}
	\end{equation*}
These residuals are expected to follow the standard normal distribution if the model is correct. Hence, the validity of a DW model can be assessed using  goodness-of-fit investigations of the normality of the residuals, such as Q-Q plots and normality tests.

\section{Assessing the performance of DW regression models} \label{sec:simulation}

This section performs a comparison of DW regression models with existing parametric approaches and with the jittering approach of \cite{machado05}. As the jittering approach fits each conditional quantile separately, we measure the performance of the models for three selected quantiles, namely $\tau=(0.25; 0.5; 0.75)$. For each $\tau$ and for each model, we evaluate the accuracy in the estimation of the conditional quantile by calculating the root mean squared error:
\begin{equation}
\textrm{RMSE} = \sqrt{\frac{\sum_{i=1}^{n}\left( \hat{\mu}_{i(\tau)} - \mu_{i(\tau)}\right) ^2}{n}}, \label{eq:RMSE}
\end{equation}
where $\mu_{i(\tau)}$ is the true quantile and $\hat{\mu}_{i(\tau)}$ is the fitted quantile from the specified model. For DW, this is calculated using \autoref{eq:quantileDW}.

We compare our approach with the following parametric approaches: Poisson, Negative Binomial, COM-Poisson, Generalised Poisson and Poisson-Inverse Gaussian. All distributions, and corresponding generalized additive models, are implemented in the R package {\tt gamlss}, with the exception of COM-Poisson, for which we use the \texttt{glm.cmp} function in the {\tt R} package {\tt COMPoissonReg} \citep{sellerspackage}. For the non-parametric jittering approach, we use the \texttt{rq.counts} function in the \texttt{Qtools} package \citep{geraci17}. Across the simulations and the different models, we use generalized additive models of the same complexity for a fair comparison. Note however that the jittering approach contains many more parameters than the parametric approaches since regression models are fitted for each quantile.

\subsection*{Simulation 1: Simulating data from our DW model}
We first simulate data from our proposed model (\autoref{eq:GAMmodelDW}). We consider different sample sizes, $n=50,100,1000$, a covariate $X \sim \textrm{Uniform}(0,1)$ and a conditional DW distribution where we link the parameters to the covariates using models of different complexity and with different associated levels of dispersion. In particular, we select four representative cases to reflect those that were considered in Figure \ref{fig:qplotDW}:
\begin{itemize}
\item{CASE 1. Linear model for $q(x)$, $\beta$ constant.}
\begin{eqnarray*}
\log\left(- \log \left(q(x) \right) \right) &=& - 5 - 3 x,  \\
\textrm{(a)} \quad \log\left(\beta\right) &=&  0.9 \quad \rightarrow \quad \textrm{over-dispersed} \\
\textrm{(b)} \quad \log\left(\beta \right) &=&  1.6 \quad \rightarrow \quad \textrm{under-dispersed}
\label{eq:case1}
\end{eqnarray*}
\item{CASE 2. Linear model for $q(x)$ and $\beta(x)$.}
\begin{eqnarray*}
\log\left(-\log \left(q(x) \right) \right) &=& - 5 - 3 x \\
\textrm{(a)} \quad \log\left(\beta (x) \right) &=&  0.6 + 0.3 x \quad \rightarrow \quad \textrm{over-dispersed} \\
\textrm{(b)} \quad \log\left(\beta (x) \right) &=&  1.1 + 0.5 x \quad \rightarrow \quad \textrm{under-dispersed}
\label{eq:case2}
\end{eqnarray*}
\item{CASE 3. Cubic B-spline model for $q(x)$, $\beta$ constant.}
\begin{eqnarray*}
\log\left(-\log \left(q(x) \right) \right) &=& - 5 - 5 x - 6 x^2 - 4 x^3 - 8 (x-g_{1})^3 I(x>g_1) +\\
&-& 9 (x-g_{2})^3 I(x>g_2) - 8 (x-g_{3})^3 I(x>g_3), \\
\textrm{(a)} \quad \log\left(\beta \right) &=&  0.9 \quad \rightarrow \quad \textrm{over-dispersed} \\
\textrm{(b)} \quad \log\left(\beta \right) &=&  1.6 \quad \rightarrow \quad \textrm{under-dispersed}
\label{eq:case3}
\end{eqnarray*}
\item{CASE 4. Cubic B-spline model for $q(x)$ and $\beta(x)$.}
\begin{eqnarray*}
\log\left(- \log \left(q(x) \right) \right) &=& - 5 - 5 x - 6 x^2 - 4 x^3 - 8 (x-g_{1})^3 I(x>g_1) -\\
&-& 9 (x-g_{2})^3 I(x>g_2) - 8 (x-g_{3})^3 I(x>g_3), \\
\textrm{(a)} \quad \log\left(\beta(x)\right) &=&  0.9 + 0.7 x + 0.9 x^2 + 0.8 x^3 + 0.9 (x-g_{1})^3 I(x>g_1) +\\
&+&  (x-g_{2})^3 I(x>g_2) + 0.9 (x-g_{3})^3 I(x>g_3) \quad \rightarrow \quad \textrm{over-dispersed} \\
\textrm{(b)} \quad \log\left(\beta(x) \right) &=&  1.6 + 1.3 x + 1.5 x^2 + 1.6 x^3 + 1.6 (x-g_{1})^3 I(x>g_1) +\\
&+& 1.6 (x-g_{2})^3 I(x>g_2) + 1.6 (x-g_{3})^3 I(x>g_3) \quad \rightarrow \quad \textrm{under-dispersed}
\label{eq:case4}
\end{eqnarray*}
\end{itemize}
Setting the values as above leads to over-dispersion values between 1.3 and 5 and under-dispersion values between 0.2 and 0.6. 

\autoref{tab:RQSD.DW_ovd} and \autoref{tab:RQSD.DW_und} report the errors in \autoref{eq:RMSE}, averaged over 100 iterations, for the three different quantiles $\tau=0.25,0.5,0.75$ and for sample sizes $n=50,100,1000$, for the over-dispersed and under-dispersed cases, respectively.
\begin{table}
\caption{\label{tab:RQSD.DW_ovd} Comparison of different models in terms of root mean squared error on over-dispersed data simulated from a DW model under four different model specifications: (1) linear link on $q(x)$, constant $\beta$, (2) linear link on both $q(x)$ and $\beta(x)$, (3) cubic B-spline on $q(x)$, constant $\beta$, (4) cubic B-spline on $q(x)$ and $\beta(x)$.}
\centering
\resizebox{\columnwidth}{!}{
\begin{tabular}{|l|c|c|c||c|c|c||c|c|c||c|c|c||c|c|c||c|c|c|} \hline
 & \multicolumn{3}{c||}{Discrete} & \multicolumn{3}{c||}{Poisson} & \multicolumn{3}{c||}{Poisson-Inverse} & \multicolumn{3}{c||}{COM-Poisson} & \multicolumn{3}{c|}{Negative} &\multicolumn{3}{c||}{Jittering}\\
& \multicolumn{3}{c||}{ Weibull} & \multicolumn{3}{c||}{} & \multicolumn{3}{c||}{Gaussian} & \multicolumn{3}{c||}{} & \multicolumn{3}{c|}{ Binomial} & \multicolumn{3}{c||}{}\\ \hline
    $\tau\backslash n$  & 50    & 100   & 1000  & 50    & 100   & 1000  & 50    & 100   & 1000  & 50    & 100   & 1000  & 50    & 100   & 1000  & 50    & 100   & 1000 \\ \hline
    (1)   &       &       &       &       &       &       &       &       &       &       &       &       &       &       &       &       &       &  \\
    0.25  & 0.990 & 0.770 & 0.423 & 2.144 & 2.120 & 2.111 & 1.062 & 0.814 & 0.490 & 1.210 & 1.034 & 0.877 & 1.040 & 0.814 & 0.515 & 1.254 & 1.064 & 0.509 \\
    0.5   & 1.126 & 0.877 & 0.443 & 1.162 & 0.926 & 0.511 & 1.268 & 1.039 & 0.783 & 1.182 & 0.920 & 0.547 & 1.197 & 0.945 & 0.635 & 1.368 & 1.023 & 0.498 \\
    0.75  & 1.388 & 1.068 & 0.483 & 2.197 & 1.992 & 1.863 & 1.607 & 1.242 & 0.866 & 1.586 & 1.214 & 0.752 & 1.509 & 1.130 & 0.664 & 1.645 & 1.200 & 0.536 \\ \hline
    (2)   &       &       &       &       &       &       &       &       &       &       &       &       &       &       &       &       &       &  \\
    0.25  & 1.784 & 1.223 & 0.604 & 4.198 & 4.036 & 4.013 & 1.807 & 1.259 & 0.688 & 1.847 & 1.246 & 0.577 & 1.771 & 1.224 & 0.634 & 2.098 & 1.414 & 0.675 \\
    0.5   & 1.870 & 1.352 & 0.686 & 1.864 & 1.418 & 0.991 & 2.150 & 1.741 & 1.348 & 1.932 & 1.335 & 0.617 & 1.906 & 1.423 & 0.917 & 2.111 & 1.557 & 0.744 \\
    0.75  & 2.253 & 1.621 & 0.810 & 3.603 & 3.453 & 3.268 & 2.423 & 1.845 & 1.259 & 2.334 & 1.594 & 0.732 & 2.304 & 1.658 & 0.938 & 2.631 & 1.941 & 0.864 \\ \hline
    (3)   &       &       &       &       &       &       &       &       &       &       &       &       &       &       &       &       &       &  \\
    0.25  & 1.764 & 1.237 & 0.506 & 2.862 & 2.652 & 2.190 & 1.972 & 1.548 & 0.703 & 2.319 & 1.983 & 1.317 & 1.953 & 1.536 & 0.753 & 2.807 & 2.159 & 0.778 \\
    0.5   & 2.183 & 1.564 & 0.609 & 2.246 & 1.742 & 0.678 & 2.295 & 1.749 & 0.890 & 2.261 & 1.752 & 0.750 & 2.228 & 1.686 & 0.766 & 2.880 & 2.096 & 0.781 \\
    0.75  & 2.727 & 1.968 & 0.710 & 3.062 & 2.471 & 1.945 & 2.915 & 2.200 & 1.136 & 2.785 & 2.128 & 1.149 & 2.808 & 2.097 & 0.911 & 3.142 & 2.336 & 0.845 \\ \hline
    (4)   &       &       &       &       &       &       &       &       &       &       &       &       &       &       &       &       &       &  \\
    0.25  & 2.586 & 1.510 & 0.828 & 2.800 & 2.490 & 2.143 & 4.752 & 1.606 & 0.808 & 4.171 & 1.997 & 0.898 & 3.674 & 1.562 & 0.790 & 2.668 & 1.237 & 0.910 \\
    0.5   & 2.370 & 1.131 & 0.704 & 2.585 & 1.118 & 0.823 & 5.299 & 1.030 & 0.720 & 4.113 & 1.836 & 0.743 & 3.076 & 0.975 & 0.562 & 4.142 & 2.088 & 0.768 \\
    0.75  & 2.464 & 1.761 & 0.669 & 3.901 & 2.000 & 1.595 & 7.595 & 1.609 & 0.723 & 4.986 & 2.272 & 0.707 & 4.072 & 1.609 & 0.534 & 2.573 & 2.285 & 0.985 \\  \hline
    \end{tabular}}
  \end{table}%
\begin{table}
\caption{\label{tab:RQSD.DW_und}Comparison of different models in terms of root mean squared error on under-dispersed data simulated from a DW model under four different model specifications: (1) linear link on $q(x)$, constant $\beta$, (2) linear link on both $q(x)$ and $\beta(x)$, (3) cubic B-spline on $q(x)$, constant $\beta$, (4) cubic B-spline on $q(x)$ and $\beta(x)$.}
\centering
\resizebox{\columnwidth}{!}{
\begin{tabular}{|l|c|c|c||c|c|c||c|c|c||c|c|c||c|c|c|} \hline
    &  \multicolumn{3}{c||}{Discrete}&  \multicolumn{3}{c||}{Poisson} & \multicolumn{3}{c||}{COM-Poisson} & \multicolumn{3}{c|}{Generalized} &  \multicolumn{3}{c||}{Jittering}\\
  &\multicolumn{3}{c||}{ Weibull} & \multicolumn{3}{c||}{} & \multicolumn{3}{c||}{} & \multicolumn{3}{c||}{Poisson} & \multicolumn{3}{c||}{}\\ \hline
  $\tau\backslash n$  & 50    & 100   & 1000  & 50    & 100   & 1000  & 50    & 100   & 1000  & 50    & 100   & 1000  & 50    & 100   & 1000 \\ \hline
    (1)   &       &       &       &       &       &       &       &       &       &       &       &       &       &       &  \\
    0.25  & 0.293 & 0.224 & 0.109 & 0.878 & 0.867 & 0.822 & 0.337 & 0.261 & 0.162 & 0.875 & 0.870 & 0.863 & 0.364 & 0.270 & 0.210 \\
    0.5   & 0.348 & 0.255 & 0.153 & 0.483 & 0.464 & 0.457 & 0.375 & 0.313 & 0.280 & 0.462 & 0.455 & 0.453 & 0.372 & 0.280 & 0.172 \\
    0.75  & 0.401 & 0.325 & 0.191 & 0.679 & 0.647 & 0.639 & 0.413 & 0.355 & 0.266 & 0.640 & 0.632 & 0.622 & 0.397 & 0.341 & 0.275 \\ \hline
    (2)   &       &       &       &       &       &       &       &       &       &       &       &       &       &       &  \\
    0.25  & 0.107 & 0.047 & 0.007 & 0.141 & 0.052 & 0.051 & 0.459 & 0.410 & 0.074 & 0.152 & 0.104 & 0.081 & 0.341 & 0.210 & 0.153 \\
    0.5   & 0.136 & 0.060 & 0.005 & 0.166 & 0.097 & 0.076 & 0.450 & 0.314 & 0.101 & 0.164 & 0.114 & 0.106 & 0.480 & 0.285 & 0.095 \\
    0.75  & 0.224 & 0.114 & 0.079 & 0.569 & 0.893 & 0.563 & 0.400 & 0.372 & 0.023 & 0.894 & 0.572 & 0.562 & 0.561 & 0.298 & 0.221 \\ \hline
    (3)   &       &       &       &       &       &       &       &       &       &       &       &       &       &       &  \\
    0.25  & 0.401 & 0.343 & 0.161 & 0.989 & 0.944 & 0.904 & 0.473 & 0.441 & 0.273 & 1.044 & 0.945 & 0.794 & 0.539 & 0.483 & 0.230 \\
    0.5   & 0.386 & 0.351 & 0.192 & 0.402 & 0.393 & 0.371 & 0.381 & 0.365 & 0.245 & 0.444 & 0.393 & 0.371 & 0.421 & 0.358 & 0.221 \\
    0.75  & 0.545 & 0.480 & 0.318 & 0.707 & 0.685 & 0.610 & 0.553 & 0.494 & 0.328 & 0.837 & 0.686 & 0.710 & 0.565 & 0.516 & 0.379 \\ \hline
    (4)   &       &       &       &       &       &       &       &       &       &       &       &       &       &       &  \\
    0.25  & 0.529 & 0.245 & 0.170 & 1.010 & 0.899 & 0.883 & 0.583 & 0.346 & 0.270 & 0.980 & 0.899 & 0.883 & 0.529 & 0.207 & 0.141 \\
    0.5   & 0.447 & 0.332 & 0.205 & 0.616 & 0.539 & 0.454 & 0.600 & 0.424 & 0.300 & 0.616 & 0.539 & 0.454 & 0.663 & 0.332 & 0.290 \\
    0.75  & 0.600 & 0.447 & 0.414 & 0.735 & 0.691 & 0.648 & 0.693 & 0.447 & 0.421 & 0.735 & 0.649 & 0.564 & 0.748 & 0.490 & 0.424 \\ \hline
    \end{tabular}}
\end{table}
Considering the case of over-dispersed data, for every $\tau$ and independently on the sample size, the Discrete Weibull outperforms the other models, followed closely by negative Binomial and the jittering approach. For the more complex case (CASE 4), jittering clearly shows greater flexibility compared to the miss-specified parametric approaches. A similar picture is given by the under-dispersed case, where discrete Weibull is followed closely by COM-Poisson and jittering. 
On the other hand, \autoref{tab:simtime} shows a clear computational gain of DW compared with COM-Poisson and jittering. The time is reported only for one simulation and, in the case of jittering, for the median and using 50 dithered samples.
\begin{table}
\caption{\label{tab:simtime} Comparison of system time (secs) for one simulation from a DW model under four different model specifications: (1) linear link on $q(x)$, constant $\beta$, (2) linear link on both $q(x)$ and $\beta(x)$, (3) cubic B-spline on $q(x)$, constant $\beta$, (4) cubic B-spline on $q(x)$ and $\beta(x)$.}
\centering
\resizebox{.6\columnwidth}{!}{%
    \begin{tabular}{|l|cccc|} \hline
    over-disp. & CASE 1 & CASE 2 & CASE 3 & CASE 4 \\ \hline
    Discrete Weibull    & 0.92  & 0.39  & 0.39  & 0.4 \\
    Poisson    & 0.05  & 0.06  & 0.04  & 0.03 \\
    Negative Binomial    & 2.05  & 0.17  & 0.12  & 0.14 \\
    Poisson-Inverse Gamma   & 0.56  & 0.34  & 0.22  & 0.23 \\
    COM-Poisson   & 6.36  & 16.18 & 33.19 & 146.78 \\
    Jittering  & 1.03  & 1.13  & 1.16  & 0.97 \\ \hline
    under-disp. & CASE 1 & CASE 2 & CASE 3 & CASE 4 \\ \hline
    Discrete Weibull    & 0.39  & 0.43  & 0.39  & 0.45 \\
    Poisson    & 0.04  & 0.01  & 0.05  & 0.04 \\
    COM-Poisson   & 5.22  & 11.34 & 68.42 & 148.36 \\
    Generalized Poisson   & 0.35  & 0.29  & 0.47  & 0.44 \\
    Jittering  & 0.87  & 0.89  & 1.06  & 1.13 \\ \hline
    \end{tabular}%
    }
\end{table}%

\subsection*{Simulation 2: Simulating data from an NB model with tail effects}

In a second simulation, we test the performance of our approach in the case of miss-specification and tail behaviour. In particular, we simulate data with a negative Binomial conditional distribution, with the parameters $\mu$ (mean) and $\sigma$ (dispersion) linked to the covariates by:
\begin{align*}
\begin{split}
\log\left(\mu(x) \right) &= 0.3 + 0.7 x_1 \\
\log\left(\sigma(x) \right) &=  -2 + 2 x_2.
\end{split}
\end{align*}
We simulate tail behaviour by letting the dispersion parameter depend on a regressor that does not affect the mean.

\autoref{tab:RQSD.NB} reports the square root of the error in \autoref{eq:RMSE}, averaged over 100 iterations, with $X_1$ and $X_2$ drawn from a $\textrm{Uniform}(0,1)$ distribution and for the different sample sizes $n=50,100,1000$.
\begin{table}
   \caption{NB data: RMSE comparison of Jittering, Discrete Weibull and Negative Binomial model.}
   \centering
  \resizebox{0.8\columnwidth}{!}{%
    \begin{tabular}{|l|r|r|r||r|r|r||r|r|r|} \hline
    NB data  & \multicolumn{3}{c||}{Jittering} & \multicolumn{3}{c||}{DW} & \multicolumn{3}{c|}{NB} \\ \hline
    \!&\! n=50  \!&\! n=100 \!&\! n=1000 \!&\! n=50  \!&\! n=100 \!&\! n=1000 \!&\! n=50  \!&\! n=100 \!&\! n=1000 \\ \hline
    $\tau$=.25 \!&\! 0.517 \!&\! 0.505 \!&\! 0.108 \!&\! 0.497 \!&\! 0.477 \!&\! 0.102 \!&\! 0.509 \!&\! 0.480 \!&\! 0.094 \\
    $\tau$=.5 \!&\! 0.538 \!&\! 0.495 \!&\! 0.095 \!&\! 0.514 \!&\! 0.479 \!&\! 0.088 \!&\! 0.555 \!&\! 0.486 \!&\! 0.084 \\
    $\tau$=.75 \!&\! 0.694 \!&\! 0.607 \!&\! 0.143 \!&\! 0.622 \!&\! 0.577 \!&\! 0.122 \!&\! 0.667 \!&\! 0.578 \!&\! 0.119 \\ \hline
    \end{tabular}%
    }
  \label{tab:RQSD.NB}%
\end{table}
The model estimates are presented in \autoref{tab:estcase2B_DW} for the Discrete Weibull, Negative Binomial and Jittering model, respectively. It is interesting to note that: (1) The Discrete Weibull model on $q(x)$ and $\beta(x)$ is behaving similarly to the Negative Binomial model, by selecting only $X_1$ significant in predicting $q(x)$, and only $X_2$ for $\beta(x)$, suggesting that the regression on both parameters helps identifying more complex dependencies such as tail behaviour. (2) The jittering approach is able to detect $\tau$-dependent significant variables, which is clearly not possible for a parametric model.
\begin{table}
  \caption{Parameter estimates for DW, NB and jittering model from NB simulated data with tail behaviour.}
  \begin{threeparttable}
    \begin{tabular}{|l|ll|ll|lll|} \hline
          & \multicolumn{2}{c|}{DW} &\multicolumn{2}{c|}{NB} & \multicolumn{3}{c|}{Jittering} \\  \hline
          & $q(x)$& $\beta(x)$  & $\mu(x)$ & $\sigma(x)$ & $\tau$=.25 &$\tau$=.5 & $\tau$=.75\\ \hline
    (Intercept) \!&\! 0.775*** \!&\! 0.564*** \!&\! 0.354*** \!&\! -2.219*** \!&\! -0.136 \!&\! 0.21* \!&\! 0.636*** \\
          \!&\! (0.062) \!&\! (0.075) \!&\! (0.078) \!&\! (0.392) \!&\! (0.135) \!&\! (0.094) \!&\! (0.103) \\
    x1    \!&\! 0.538*** \!&\! -0.116 \!&\! 0.696*** \!&\! 0.764. \!&\! 0.434* \!&\! 0.678*** \!&\! 0.802*** \\
          \!&\! (0.084) \!&\! (0.096) \!&\! (0.101) \!&\! (0.41) \!&\! (0.185) \!&\! (0.131) \!&\! (0.133) \\
    x2    \!&\! -0.094 \!&\! -0.364*** \!&\! -0.032 \!&\! 1.466*** \!&\! -0.503* \!&\! -0.219 \!&\! -0.026 \\
          \!&\! (0.088) \!&\! (0.096) \!&\! (0.103) \!&\! (0.428) \!&\! (0.201) \!&\! (0.146) \!&\! (0.158) \\ \hline
    \end{tabular}%
  \label{tab:estcase2B_DW}%
    \begin{tablenotes}
\footnotesize
\item Signif. codes:  0 '***', 0.001 '**', 0.01 '*', 0.05 '.'
\end{tablenotes}
  \end{threeparttable}
\end{table}%

\section{Modelling the relationship between family background and planned fertility} \label{sec:realdata}
\begin{figure}
\centering
\includegraphics[scale=0.4]{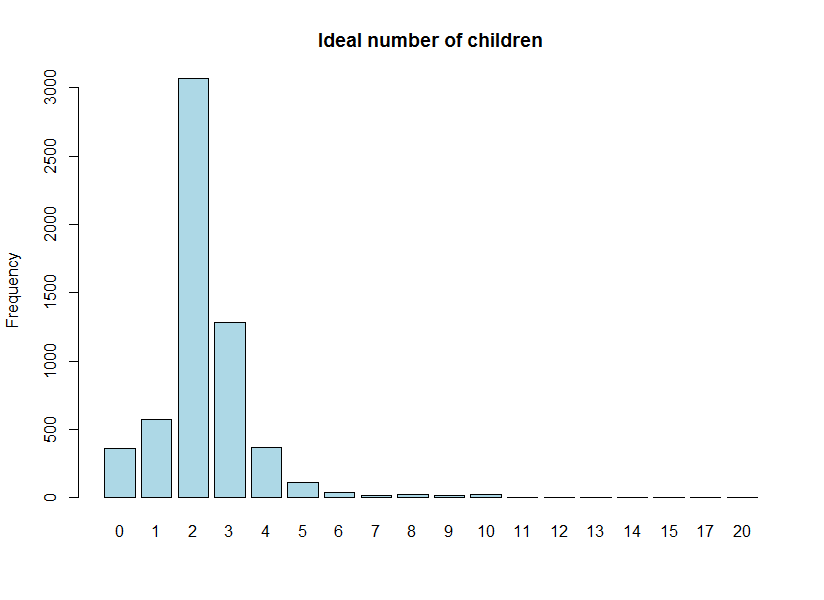}
\caption{Distribution of the response variable: ideal number of children for 15-17 years old Mexican women.}
     	\label{fig:response}
\end{figure}

We use the latest data from the ENADID National Survey of Demographic Dynamics in Mexico \citep{inegi14} to study the effect of education and family background on fertility plans. In particular, we take as dependent variable the number of planned children declared by young Mexican women at the ENADID interview. As in \cite{miranda08}, we consider women between 15 and 17 years old who at the time of the interview were living with at least one biological parent and had neither started independent economic life nor entered motherhood. This selection avoids any confusion between current and planned fertility and ensures that all individuals are broadly at the same point of their life-cycle. A number of covariates are selected to control for education and family background: whether the teenager can speak an indigenous/native language, whether primary, secondary and higher education attainments were completed, a wealth index (low, medium low, medium high, and high), the location of the parental household (rural, urban, and suburban) and a set of variables describing the socio-economic characteristics of the head of the family, amongst which the age and gender and the same education attainment covariates considered for the teenagers. Finally, the total number of persons living with the teenager (family size) and a series of dummies indicating the state of residence (32 in total) are also used as explanatory variables. This gives a total of 53 explanatory variables in this study and a sample size of 5906 women. \autoref{fig:response} shows the distribution of the response variable. Without taking into consideration the effect of the covariates, the distribution of ideal fertility shows under-dispersion relative to Poisson with a dispersion index of 0.86, higher than that of the 1997 data (0.55), possibly due to a small number of outliers (8 women) declaring more than 12 planned children (the maximum in \cite{miranda08}).

\begin{figure}
\centering
 \begin{minipage}[b]{0.45\textwidth}
    \includegraphics[width=\textwidth]{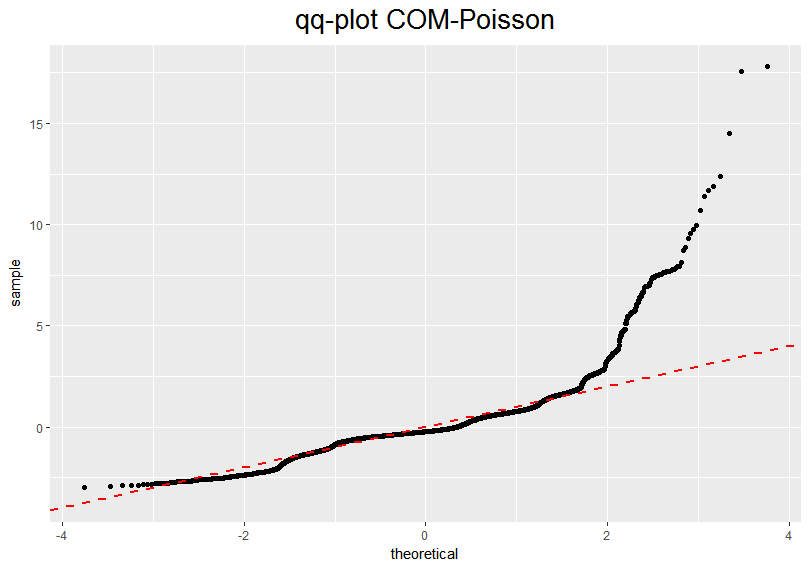}
  \end{minipage}
 \hfill
  \begin{minipage}[b]{0.45\textwidth}
    \includegraphics[width=\textwidth]{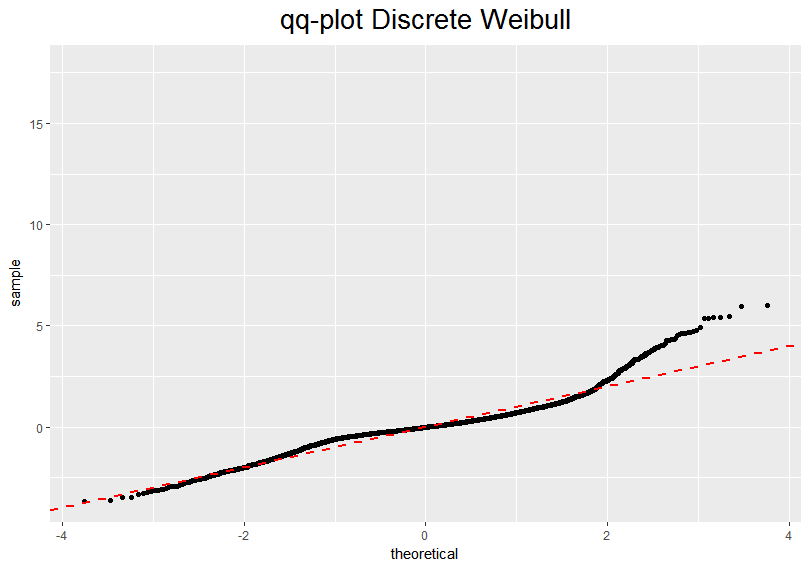}
  \end{minipage}
  \caption{Q-Q plot of the randomized quantile residuals of the DW and COM-Poisson generalized additive models fitted to the ideal fertility data.}
\label{fig:qplotDW-fertility}
\end{figure}

We fit Poisson, COM-Poisson and Discrete Weibull generalized additive models to these data (generalized Poisson had problems of convergence on this dataset). Given the computational complexity of COM-Poisson but in the interest of a fair comparison, we select the best model with the following strategy: we fix a linear link on all parameters (one for Poisson and two for DW and COM-Poisson), then we consider the inclusion of possible non-linear terms for the two non-categorical variables (age of head of the family and family size) in a forward stepwise manner. For this, we select the same level of complexity for each parameter and we search for all model combinations up to a maximum of degree three of the polynomial and three internal knots (cubic spline). The best model at each step is selected based on AIC. \autoref{tab:SMTHidealf} shows the complexity of the selected model and \autoref{fig:qplotDW-fertility} shows the fitting of the top two models in terms of randomized quantile residuals. Overall, DW and COM-Poisson appear to provide a similar fit to the data, with a slightly lower AIC for COM-Poisson but a worse fit of the Gaussian distribution to the residuals. The computational time, here reported only for the best model in the search, shows a striking difference between DW and COM-Poisson, which limits further comparisons and more extensive searches.
\begin{table}
  \caption{The Poisson, Discrete Weibull and COM-Poisson generalized additive models selected on the ideal fertility data with a B-spline (with degree D and number of internal knots $k$) for two of the variables. The last column reports the system time (in seconds).}
  \centering
  \resizebox{0.9\columnwidth}{!}{%
    \begin{tabular}{|l|c|c|c|cc|cc|c|} \hline
&  AIC   & LogLik & \# param & \multicolumn{2}{|c|}{age (family head)} & \multicolumn{2}{|c|}{family size} & Time (secs)\\
                 &       &       &       & D     & k     & D     & k & \\ \hline
    PO      & 19779.59 & -9838.796 & 51    & 1     & 1     & 1     & 0 & 0.14\\
    DW    & 19351.46 & -9561.731 & 114   & 2     & 3     & 2     & 2   & 9.67 \\
    CMP    & 19175.24 & -9475.621 & 112   & 1     & 1     & 3     & 3 & 7339.46\\ \hline
    \end{tabular}}
  \label{tab:SMTHidealf}%
\end{table}%

\begin{figure}
\centering
 \begin{minipage}[b]{0.49\textwidth}
   \includegraphics[width=\textwidth]{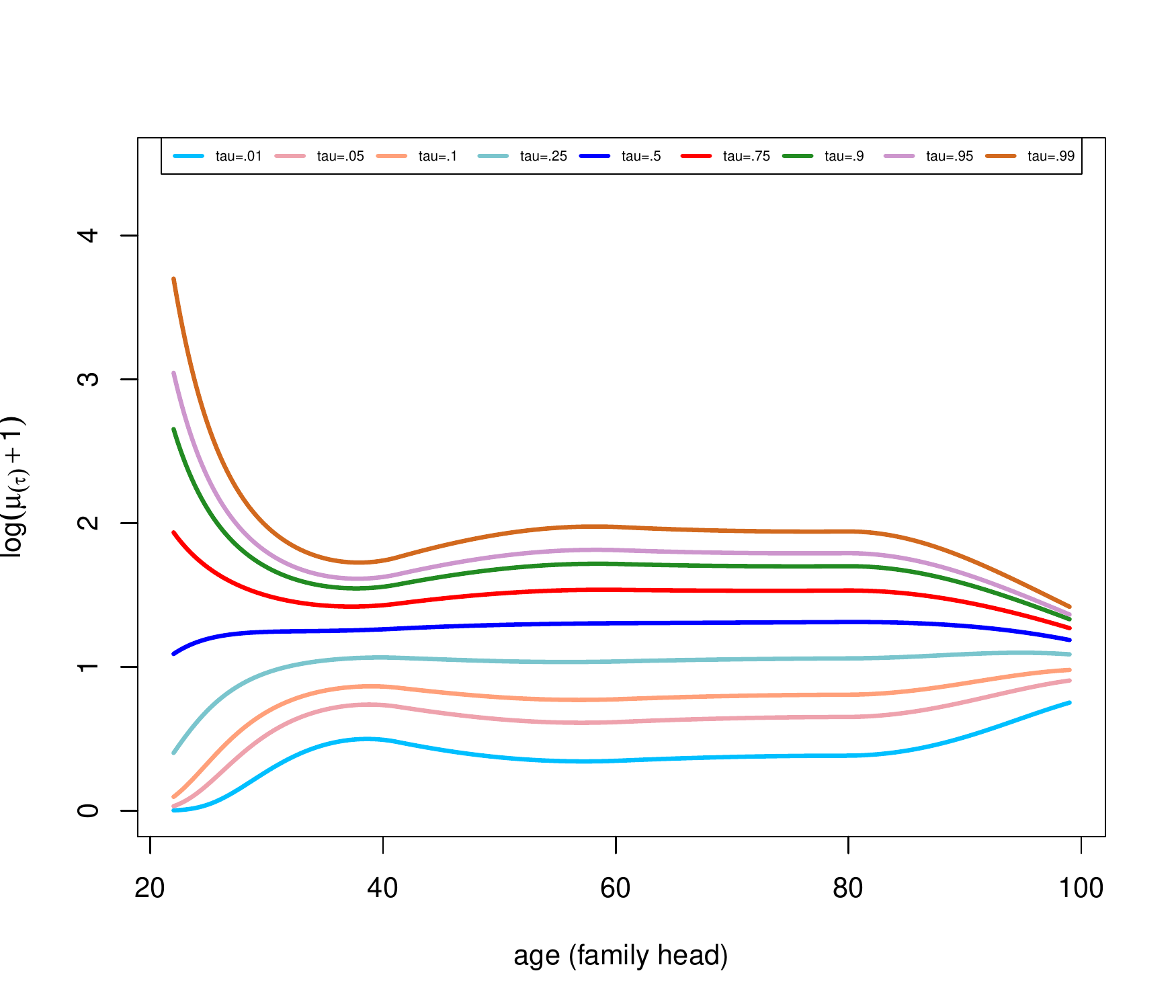} \\  \centering age (family head) - DW
  \end{minipage}
 \hfill
  \begin{minipage}[b]{0.49\textwidth}
    \includegraphics[width=\textwidth]{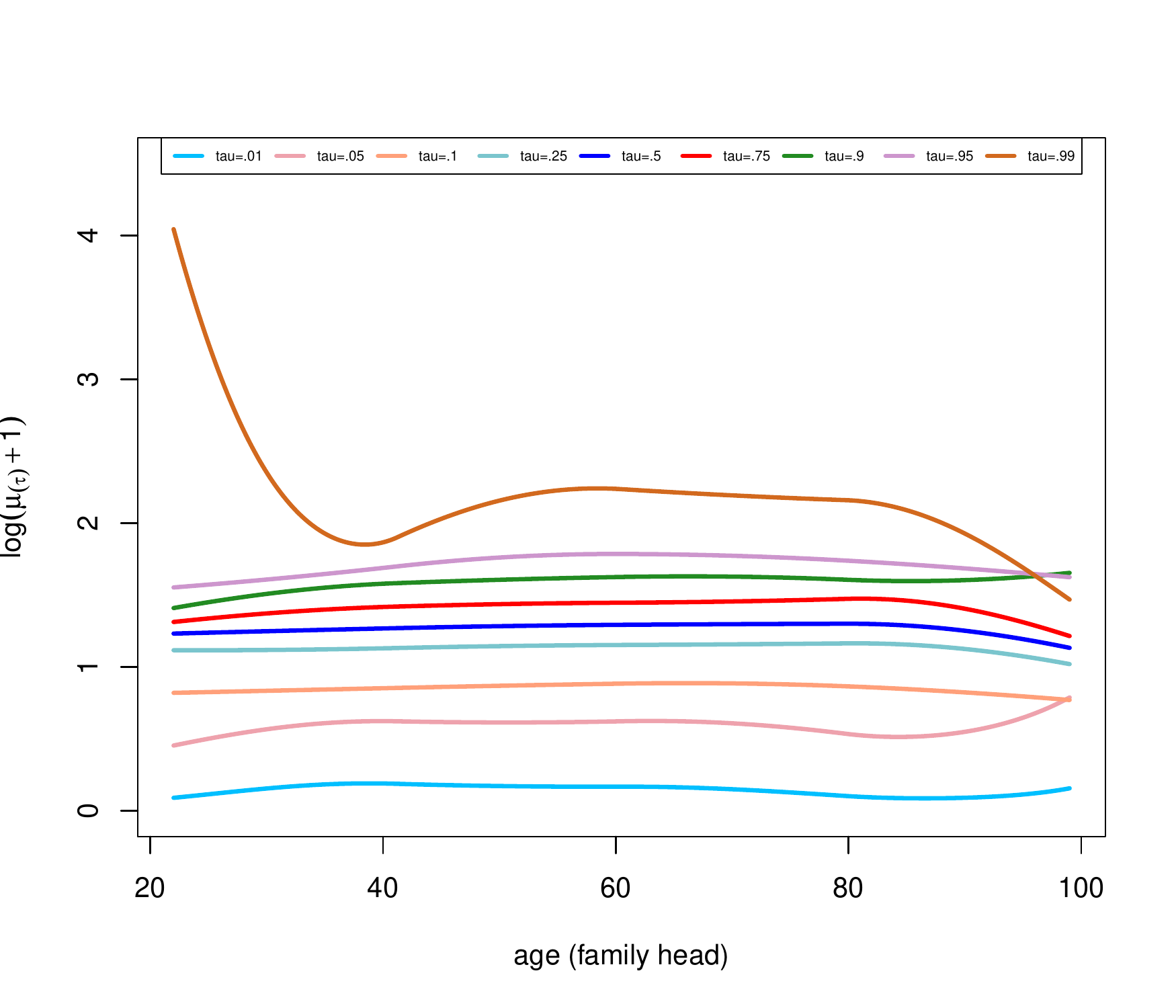} \\   \centering age (family head) - Jittering
  \end{minipage}
 \hfill
  \begin{minipage}[b]{0.49\textwidth}
    \includegraphics[width=\textwidth]{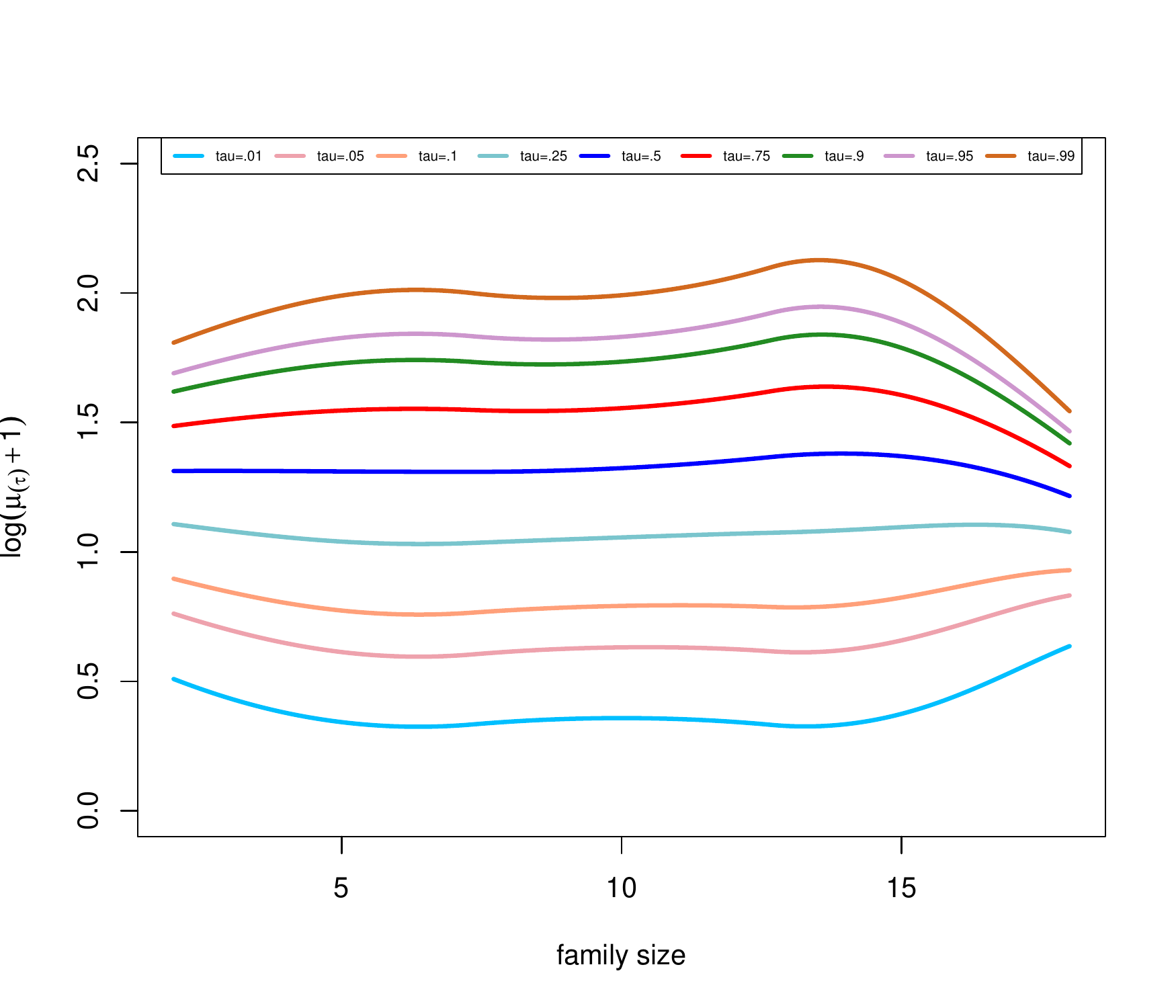} \\   \centering family size - DW
  \end{minipage}
 \hfill
  \begin{minipage}[b]{0.49\textwidth}
    \includegraphics[width=\textwidth]{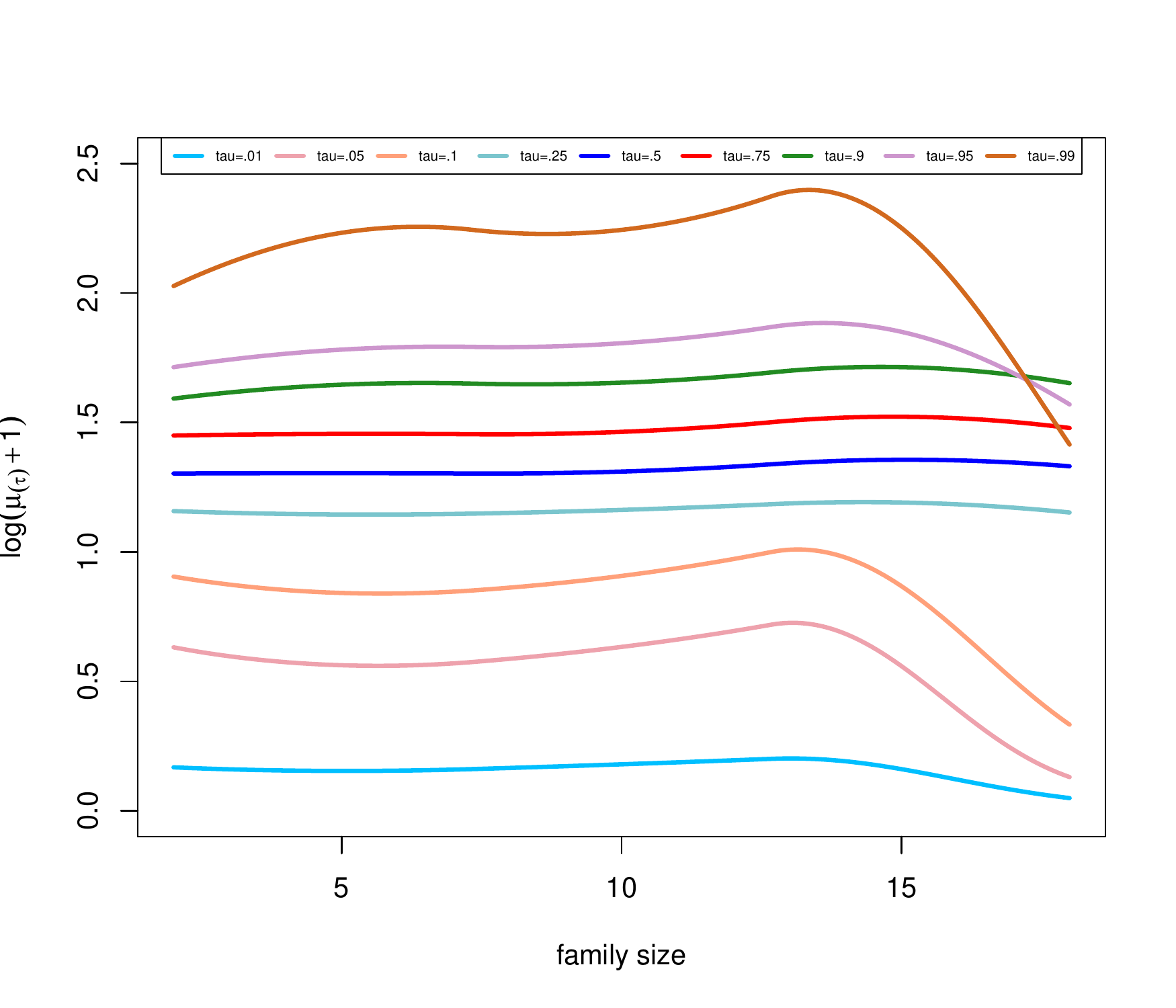}\\   \centering family size - Jittering
  \end{minipage}
  \caption{Plot of the conditional $\tau$-quantiles of ideal fertility for two variables, age of the family head (top) and family size (bottom), keeping all the other covariates fixed to their mean, for the DW generalized additive model (left) and the jittering approach (right).}
\label{fig:CQidealf}
\end{figure}

The use of a generalized additive model resulted in a quadratic spline for the two continuous variables in this study. In an attempt to measure how well a flexible parametric approach can approximate the conditional distribution of ideal fertility given the explanatory variables, we compare the partial effects obtained from our model with those of a more flexible non-parametric jittering approach where a quadratic spline (with the same number of internal knots as in the DW model) is fitted to each conditional quantile, thus resulting in a larger number of parameters. \autoref{tab:PEidealf} shows similar levels of effects for all the variables, both in terms of sign and intensity. Both approaches are able to capture tail effects in the distribution, with a number of variables exhibiting sign reversals of the effects. The conclusions are similar to those obtained by \cite{miranda08} in their earlier study, with variables related to education and family background being highly significant. In our analysis, and also using COM-Poisson, the education indicators of the head of the family appear to be highly significant whereas these are not picked up as significant by the jittering approach although the marginal effects are close. This may be down to a higher instability in the estimation of the standard errors for jittering, possibly due to the larger number of parameters and also to the uniform random sampling underlying the method (100 samples are used for the results presented in this paper). A further example of this is with the variable family size, which is not found significant for COM-Poisson (p-values above 0.1), but is found highly significant for the DW model and only for some of the quantiles for the jittering approach (e.g. 0.8 and 0.99 but not 0.9 and 0.95). This may  limit the interpretation of the results of the jittering approach, making the conclusions overly sensitive to the specific quantile selected. \autoref{fig:CQidealf} shows further how this discontinuity could be the result of crossing of quantiles, which is the drawback of non-parametric approaches that fit models individually for each quantile. Although the general trends are similar between DW and jittering, the jittering approach produces crossing of quantiles at the extreme of the distribution, where there is usually a small sample size.

\begin{table}
\caption{Partial effects of the regressors on ideal fertility for the jittering (top) and the discrete Weibull (bottom) generalized additive models. Significant variables at the 5\% level are highlighted in bold.}
\centering
\resizebox{\columnwidth}{!}{%
    \begin{tabular}{|l|rrrrrrrrrrrrr|} \hline
    $\tau$ & \textit{0.01} & \textit{0.05} & \textit{0.1} & \textit{0.25} & \textit{0.3} & \textit{0.4} & \textit{0.5} & \textit{0.6} & \textit{0.75} & \textit{0.8} & \textit{0.9} & \textit{0.95} & \textit{0.99} \\ \hline
    \multicolumn{14}{|c|}{Jittering}  \\ \hline
    family size & 0.011 & -0.008 & 0.013 & -0.01 & -0.016 & \textbf{-0.025} & \textbf{-0.038} & \textbf{-0.047} & \textbf{-0.054} & \textbf{-0.049} & -0.069 & -0.076 & \textbf{-0.362} \\
    HFage & 0.019 & 0.077 & 0.038 & 0.037 & 0.042 & 0.04  & \textbf{0.035} & \textbf{0.05} & \textbf{0.041} & \textbf{0.063} & 0.12  & 0.168 & 0.317 \\
    indspeaker & \textbf{-0.099} & -0.296 & -0.257 & 0.008 & 0.018 & 0.018 & 0.044 & 0.068 & 0.198 & 0.214 & 0.138 & 0.146 & -0.379 \\
    cprimary & 1.391 & 2.956 & 2.012 & \textbf{6.971} & \textbf{5.409} & \textbf{3.675} & \textbf{2.929} & \textbf{3.172} & 2.216 & 1.488 & \textbf{1.221} & \textbf{1.624} & 5.082 \\
    isecondary & 3.755 & 5.952 & 3.561 & \textbf{7.233} & \textbf{5.572} & \textbf{3.61} & \textbf{2.729} & \textbf{2.872} & 1.942 & 1.196 & \textbf{0.89} & \textbf{1.447} & 5.443 \\
    csecondary & 4.178 & 7.119 & 4.293 & \textbf{7.336} & \textbf{5.878} & \textbf{3.851} & \textbf{2.82} & \textbf{3.124} & 2.047 & 1.172 & 0.706 & 1.837 & 3.016 \\
    osecondary & 3.34  & 6.028 & 3.814 & \textbf{7.335} & \textbf{5.727} & \textbf{3.682} & \textbf{2.811} & \textbf{2.956} & 2.019 & 1.278 & \textbf{0.985} & \textbf{1.481} & 5.549 \\
    wealth\_mlow & -0.024 & -0.132 & \textbf{-0.191} & \textbf{-0.077} & \textbf{-0.083} & \textbf{-0.088} & \textbf{-0.111} & \textbf{-0.111} & \textbf{-0.167} & \textbf{-0.189} & \textbf{-0.166} & 0.109 & 0.564 \\
    wealth\_mhigh & -0.039 & -0.187 & \textbf{-0.223} & \textbf{-0.074} & \textbf{-0.082} & \textbf{-0.105} & \textbf{-0.128} & \textbf{-0.137} & \textbf{-0.183} & \textbf{-0.169} & -0.096 & 0.245 & \textbf{2.477} \\
    wealth\_high & 0.034 & -0.127 & \textbf{-0.283} & -0.087 & -0.062 & -0.04 & -0.07 & -0.029 & -0.063 & -0.058 & 0.052 & 0.979 & \textbf{3.852} \\
    urban & -0.033 & -0.003 & -0.024 & -0.041 & \textbf{-0.046} & \textbf{-0.054} & \textbf{-0.058} & \textbf{-0.081} & \textbf{-0.098} & -0.111 & -0.05 & -0.112 & 0.02 \\
    surban & -0.02 & 0.041 & 0.028 & -0.019 & -0.011 & -0.024 & -0.04 & -0.041 & -0.046 & -0.071 & 0.008 & -0.028 & 1.197 \\
    HFmale & 0.048 & 0.104 & 0.059 & 0.006 & 0.033 & 0.019 & 0.011 & 0.03  & 0.053 & 0.041 & 0.039 & -0.044 & -1.062 \\
    HFcprimary & 0.041 & 0.196 & 0.341 & 0.111 & \textbf{0.109} & 0.073 & 0.056 & 0.087 & 0.061 & 0.041 & -0.117 & -0.218 & -1.664 \\
    HFisecondary & 0.006 & 0.118 & 0.26  & 0.101 & 0.084 & 0.039 & 0.029 & 0.043 & 0.021 & 0.006 & -0.185 & -0.338 & -1.151 \\
    HFcsecondary & -0.021 & -0.164 & -0.161 & -0.031 & -0.002 & 0.023 & 0.09  & 0.136 & 0.178 & 0.224 & -0.039 & -0.324 & -2.209 \\
    HFosecondary & -0.03 & -0.017 & 0.183 & 0.062 & 0.06  & 0.027 & 0.022 & 0.034 & 0.085 & 0.093 & 0.06  & 0.163 & 0.031 \\
    HFindspeaker & -0.004 & -0.155 & -0.21 & -0.051 & -0.06 & -0.038 & -0.04 & -0.036 & 0.005 & 0.049 & 0.05  & -0.142 & -0.756 \\ \hline
    \multicolumn{14}{|c|}{Discrete Weibull}  \\ \hline
    family size& \textbf{0.023} & \textbf{0.023} & \textbf{0.017} & \textbf{-0.003} & \textbf{-0.011} & \textbf{-0.026} & \textbf{-0.042} & \textbf{-0.061} & \textbf{-0.096} & \textbf{-0.111} & \textbf{-0.155} & \textbf{-0.194} & \textbf{-0.273} \\
    HFage & \textbf{0.004} & \textbf{0.01} & \textbf{0.016} & \textbf{0.028} & \textbf{0.031} & \textbf{0.038} & \textbf{0.045} & \textbf{0.053} & \textbf{0.066} & \textbf{0.072} & \textbf{0.087} & \textbf{0.1} & \textbf{0.126} \\
    indspeaker & -0.001 & -0.001 & -0.001 & 0.001 & 0.001 & 0.002 & 0.004 & 0.005 & 0.007 & 0.008 & 0.011 & 0.014 & 0.019 \\
    cprimary & \textbf{1.333} & \textbf{1.824} & \textbf{2.061} & \textbf{2.386} & \textbf{2.451} & \textbf{2.551} & \textbf{2.625} & \textbf{2.683} & \textbf{2.741} & \textbf{2.753} & \textbf{2.756} & \textbf{2.731} & \textbf{2.627} \\
    isecondary & \textbf{1.482} & \textbf{1.952} & \textbf{2.164} & \textbf{2.429} & \textbf{2.476} & \textbf{2.542} & \textbf{2.582} & \textbf{2.603} & \textbf{2.598} & \textbf{2.584} & \textbf{2.517} & \textbf{2.434} & \textbf{2.22} \\
    csecondary & \textbf{2.069} & \textbf{2.507} & \textbf{2.671} & \textbf{2.814} & \textbf{2.825} & \textbf{2.82} & \textbf{2.79} & \textbf{2.738} & \textbf{2.607} & \textbf{2.541} & \textbf{2.337} & \textbf{2.139} & \textbf{1.71} \\
    osecondary & \textbf{1.562} & \textbf{2.041} & \textbf{2.254} & \textbf{2.517} & \textbf{2.562} & \textbf{2.624} & \textbf{2.66} & \textbf{2.676} & \textbf{2.663} & \textbf{2.645} & \textbf{2.568} & \textbf{2.475} & \textbf{2.243} \\
    wealth\_mlow & \textbf{-0.168} & \textbf{-0.243} & \textbf{-0.268} & \textbf{-0.265} & \textbf{-0.254} & \textbf{-0.222} & \textbf{-0.179} & \textbf{-0.124} & \textbf{-0.005} & \textbf{0.052} & \textbf{0.223} & \textbf{0.387} & \textbf{0.746} \\
    wealth\_mhigh & \textbf{-0.256} & \textbf{-0.383} & \textbf{-0.424} & \textbf{-0.405} & \textbf{-0.38} & \textbf{-0.309} & \textbf{-0.215} & \textbf{-0.093} & \textbf{0.172} & \textbf{0.298} & \textbf{0.684} & \textbf{1.059} & \textbf{1.888} \\
    wealth\_high & \textbf{-0.252} & \textbf{-0.368} & \textbf{-0.397} & \textbf{-0.344} & \textbf{-0.307} & \textbf{-0.213} & \textbf{-0.094} & \textbf{0.057} & \textbf{0.377} & \textbf{0.528} & \textbf{0.984} & \textbf{1.423} & \textbf{2.387} \\
    urban & -0.002 & -0.008 & -0.013 & -0.027 & -0.031 & -0.039 & -0.047 & -0.055 & -0.071 & -0.077 & -0.096 & -0.111 & -0.142 \\
    surban & -0.036 & -0.045 & -0.044 & -0.032 & -0.026 & -0.014 & 0.001 & 0.018 & 0.052 & 0.067 & 0.111 & 0.152 & 0.237 \\
    HFmale & \textbf{0.094} & \textbf{0.117} & \textbf{0.118} & \textbf{0.095} & \textbf{0.084} & \textbf{0.06} & \textbf{0.033} & \textbf{0.000} & \textbf{-0.064} & \textbf{-0.093} & \textbf{-0.176} & \textbf{-0.252} & \textbf{-0.409} \\
    HFcprimary & \textbf{0.334} & \textbf{0.391} & \textbf{0.387} & \textbf{0.313} & \textbf{0.281} & \textbf{0.212} & \textbf{0.134} & \textbf{0.044} & \textbf{-0.129} & \textbf{-0.205} & \textbf{-0.422} & \textbf{-0.618} & \textbf{-1.017} \\
    HFisecondary & \textbf{0.234} & \textbf{0.279} & \textbf{0.277} & \textbf{0.221} & \textbf{0.197} & \textbf{0.144} & \textbf{0.083} & \textbf{0.013} & \textbf{-0.124} & \textbf{-0.184} & \textbf{-0.357} & \textbf{-0.513} & \textbf{-0.833} \\
    HFcsecondary & \textbf{0.328} & \textbf{0.373} & \textbf{0.359} & \textbf{0.266} & \textbf{0.229} & \textbf{0.151} & \textbf{0.063} & \textbf{-0.037} & \textbf{-0.227} & \textbf{-0.31} & \textbf{-0.546} & \textbf{-0.757} & \textbf{-1.184} \\
    HFosecondary & \textbf{0.102} & \textbf{0.134} & \textbf{0.142} & \textbf{0.133} & \textbf{0.127} & \textbf{0.11} & \textbf{0.09} & \textbf{0.065} & \textbf{0.014} & \textbf{-0.009} & \textbf{-0.077} & \textbf{-0.141} & \textbf{-0.274} \\
    HFindspeaker & 0.004 & -0.008 & -0.021 & -0.053 & -0.063 & -0.083 & -0.103 & -0.126 & -0.166 & -0.184 & -0.232 & -0.274 & -0.357 \\ \hline
 \end{tabular}}
 \label{tab:PEidealf}
\end{table}

\section{Conclusions}
Motivated by an investigation about the dependency of planned fertility on education and family background, we develop a novel regression model for count data based on the discrete Weibull distribution, which has had limited use to date. We show how a regression model based on this distribution can provide a simple and unified framework to capture different levels of dispersion in the data, namely under-dispersion and over-dispersion. Given the expected complex dependencies in the planned fertility study, we develop a generalized additive model to link both parameters of the distribution to the explanatory variables.

Through a simulation study and the real data application we show some important features of the proposed models, which could potentially lead to their wide applicability to the modelling of count data. Firstly, the conditional quantiles have a simple analytical formula, which makes the calculation of partial effects straightforward as well as the interpretation of the regression coefficients. Secondly, the likelihood is the same of that of a continuous Weibull distribution with interval-censored data, so efficient implementations are already available in the R package {\tt gamlss}, for a range of models, including mixed and mixture models, and inferential procedures, including penalised likelihood approaches. Thirdly, the distribution can capture both cases of over and under-dispersion, similarly to COM-Poisson but at a fraction of the time. Fourthly, a generalized additive DW model is able to compete against the more flexible quantile regression approaches, without the need of individual fitting at each quantile and without the inherent issue of conditional quantiles' crossing.

\section*{Acknowledgement}
We thank Prof Joao Santos Silva for useful discussions and insights on this paper.

\bibliographystyle{rss}	
\bibliography{references-dw}

\begin{thebibliography}{38}
\expandafter\ifx\csname natexlab\endcsname\relax\def\natexlab#1{#1}\fi
\expandafter\ifx\csname url\endcsname\relax
  \def\url#1{\texttt{#1}}\fi
\expandafter\ifx\csname urlprefix\endcsname\relax\def\urlprefix{URL: }\fi

\bibitem[{Barbiero(2015)}]{barbiero2013package}
Barbiero, A. (2015) \textit{DiscreteWeibull: Discrete Weibull Distributions
  (Type 1 and 3)}.
\newblock \urlprefix\url{http://CRAN.R-project.org/package=DiscreteWeibull}.
\newblock R package version 1.0.1.

\bibitem[{Cameron and Trivedi(2013)}]{cameron2013regression}
Cameron, A.~C. and Trivedi, P.~K. (2013) \textit{Regression analysis of count
  data}.
\newblock Cambridge University Press.

\bibitem[{Chakraborty(2015)}]{chakraborty2015generating}
Chakraborty, S. (2015) Generating discrete analogues of continuous probability
  distributions-{A} survey of methods and constructions.
\newblock \textit{Journal of Statistical Distributions and Applications},
  \textbf{2}, 1--30.

\bibitem[{Chanialidis et~al.(2017)Chanialidis, Evers, Neocleous and
  Nobile}]{chanialidis17}
Chanialidis, C., Evers, L., Neocleous, T. and Nobile, A. (2017) Efficient
  {Bayesian} inference for {COM-Poisson} regression models.
\newblock \textit{Statistics and Computing}.

\bibitem[{Consul and Famoye(1992)}]{consul92}
Consul, P. and Famoye, F. (1992) Generalized {Poisson} regression model.
\newblock \textit{Communications in Statistics-Theory and Methods},
  \textbf{21}, 89--109.

\bibitem[{De~Boor(1972)}]{de1972calculating}
De~Boor, C. (1972) On calculating with {B-splines}.
\newblock \textit{Journal of Approximation Theory}, \textbf{6}, 50--62.

\bibitem[{Dierckx(1995)}]{dierckx1995curve}
Dierckx, P. (1995) \textit{Curve and surface fitting with splines}.
\newblock Oxford University Press.

\bibitem[{Dunn and Smyth(1996)}]{dunn1996randomized}
Dunn, P.~K. and Smyth, G.~K. (1996) Randomized quantile residuals.
\newblock \textit{Journal of Computational and Graphical Statistics},
  \textbf{5}, 236--244.

\bibitem[{Englehardt et~al.(2012)Englehardt, Ashbolt, Loewenstine, Gadzinski
  and Ayenu-Prah}]{englehardt2012methods}
Englehardt, J.~D., Ashbolt, N.~J., Loewenstine, C., Gadzinski, E.~R. and
  Ayenu-Prah, A.~Y. (2012) Methods for assessing long-term mean pathogen count
  in drinking water and risk management implications.
\newblock \textit{Journal of Water and Health}, \textbf{10}, 197--208.

\bibitem[{Englehardt and Li(2011)}]{englehardt2011discrete}
Englehardt, J.~D. and Li, R. (2011) The discrete {Weibull} distribution: an
  alternative for correlated counts with confirmation for microbial counts in
  water.
\newblock \textit{Risk Analysis}, \textbf{31}, 370--381.

\bibitem[{Geraci(2017)}]{geraci17}
Geraci, M. (2017) \textit{Qtools: {U}tilities for quantiles}.
\newblock \urlprefix\url{https://cran.r-project.org/package=Qtools}.
\newblock R package version 1.2.

\bibitem[{Haselimashhadi et~al.(2017)Haselimashhadi, Vinciotti and
  Yu}]{haselimashhadi17}
Haselimashhadi, H., Vinciotti, V. and Yu, K. (2017) A novel {Bayesian}
  regression model for counts with an application to health data.
\newblock \textit{Journal of Applied Statistics}, 1--21.

\bibitem[{Hilbe(2014)}]{hilbe2014modeling}
Hilbe, J.~M. (2014) \textit{Modeling Count Data}.
\newblock Cambridge University Press.

\bibitem[{Horowitz(1992)}]{horowitz92}
Horowitz, J. (1992) A smooth binary score estimator for the binary response
  model.
\newblock \textit{Econometrica}, \textbf{60}, 505--531.

\bibitem[{{INEGI}(2014)}]{inegi14}
{INEGI} (2014) National survey of demographic dynamics 2014.

\bibitem[{Kadane et~al.(2006)Kadane, Krishnan and Shmueli}]{kadane06}
Kadane, J.~B., Krishnan, R. and Shmueli, G. (2006) A data disclosure policy for
  count data based on the {COM-Poisson} distribution.
\newblock \textit{Management Science}, \textbf{52}, 1610--1617.

\bibitem[{Kalktawi et~al.(2015)Kalktawi, Vinciotti and Yu}]{kalktawi2015simple}
Kalktawi, H.~S., Vinciotti, V. and Yu, K. (2015) A simple and adaptive
  dispersion regression model for count data.
\newblock \textit{arXiv preprint arXiv:1511.00634}.

\bibitem[{Khan et~al.(1989)Khan, Khalique and Abouammoh}]{khan1989estimating}
Khan, M.~A., Khalique, A. and Abouammoh, A. (1989) On estimating parameters in
  a discrete {Weibull} distribution.
\newblock \textit{IEEE Transactions on Reliability}, \textbf{38}, 348--350.

\bibitem[{Kimberly et~al.(2010)Kimberly, Thomas and Andrew}]{sellerspackage}
Kimberly, S., Thomas, L. and Andrew, R. (2010) \textit{{COMPoissonReg}:
  COM-Poisson and Zero-Inflated COM-Poisson regression}.
\newblock \urlprefix\url{https://CRAN.R-project.org/package=COMPoissonReg}.
\newblock R package version 0.4.1.

\bibitem[{Knodel and Prachuabmoh(1973)}]{knodel1973desired}
Knodel, J. and Prachuabmoh, V. (1973) Desired family size in {Thailand}: Are
  the responses meaningful?
\newblock \textit{Demography}, \textbf{10}, 619--637.

\bibitem[{Kulasekera(1994)}]{kulasekera1994approximate}
Kulasekera, K. (1994) Approximate {MLEs} of the parameters of a discrete
  {Weibull} distribution with type {I} censored data.
\newblock \textit{Microelectronics Reliability}, \textbf{34}, 1185--1188.

\bibitem[{Lee and Neocleus(2010)}]{lee10}
Lee, D. and Neocleus, T. (2010) Bayesian quantile regression for count data
  with application to environmental epidemiology.
\newblock \textit{Journal of the Royal Statistical Society - Series C},
  \textbf{59}, 905--920.

\bibitem[{Lee(1992)}]{lee92}
Lee, M. (1992) Median regression for ordered discrete response.
\newblock \textit{Journal of Econometrics}, \textbf{51}, 59--77.

\bibitem[{Machado and Santos~Silva(2005)}]{machado05}
Machado, J. and Santos~Silva, M. (2005) Quantiles for counts.
\newblock \textit{JASA}, \textbf{100}, 1226--1237.

\bibitem[{Manski(1985)}]{manski85}
Manski, C. (1985) Semiparametric analysis of discrete response: asymptotic
  properties of the maximum score estimator.
\newblock \textit{Journal of Econometrics}, \textbf{3}, 205--228.

\bibitem[{Miranda(2008)}]{miranda08}
Miranda, A. (2008) Planned fertility and family background: a quantile
  regression for counts analysis.
\newblock \textit{Journal of Population Economics}, \textbf{21}, 67--81.

\bibitem[{Nakagawa and Osaki(1975)}]{nakagawa1975discrete}
Nakagawa, T. and Osaki, S. (1975) The discrete {Weibull} distribution.
\newblock \textit{IEEE Transactions on Reliability}, \textbf{24}, 300--301.

\bibitem[{Nelder and Wedderburn(1972)}]{nelder1972generalized}
Nelder, J.~A. and Wedderburn, R.~W. (1972) Generalized linear models.
\newblock \textit{Journal of the Royal Statistical Society. Series A},
  \textbf{135}, 370--384.

\bibitem[{Noufaily and Jones(2013)}]{noufaily2013parametric}
Noufaily, A. and Jones, M. (2013) Parametric quantile regression based on the
  generalized gamma distribution.
\newblock \textit{Journal of the Royal Statistical Society: Series C (Applied
  Statistics)}.

\bibitem[{Pritchett(1994)}]{pritchett1994desired}
Pritchett, L.~H. (1994) Desired fertility and the impact of population
  policies.
\newblock \textit{Population and Development Review}, \textbf{20}, 1--55.

\bibitem[{Rigby and Stasinopoulos(2005)}]{rigby05}
Rigby, R.~A. and Stasinopoulos, D.~M. (2005) Generalized additive models for
  location, scale and shape.
\newblock \textit{Journal of the Royal Statistical Society: Series C (Applied
  Statistics)}, \textbf{54}, 507--554.

\bibitem[{S{\'a}ez-Castillo and Conde-S{\'a}nchez(2013)}]{saez2013hyper}
S{\'a}ez-Castillo, A. and Conde-S{\'a}nchez, A. (2013) A {hyper-Poisson}
  regression model for overdispersed and underdispersed count data.
\newblock \textit{Computational Statistics \& Data Analysis}, \textbf{61},
  148--157.

\bibitem[{Sellers and Shmueli(2010)}]{sellers2008flexible}
Sellers, K.~F. and Shmueli, G. (2010) A flexible regression model for count
  data.
\newblock \textit{Annals of Applied Statistics}, \textbf{4}, 943--961.

\bibitem[{Smith and Faddy(2016)}]{smith16}
Smith, D. and Faddy, M. (2016) Mean and variance modeling of under- and
  overdispersed count data.
\newblock \textit{Journal of Statistical Software, Articles}, \textbf{69},
  1--23.

\bibitem[{Stasinopoulos et~al.(2007)Stasinopoulos, Rigby
  et~al.}]{stasinopoulos2007generalized}
Stasinopoulos, D.~M., Rigby, R.~A. et~al. (2007) {Generalized additive models
  for location scale and shape (GAMLSS) in R}.
\newblock \textit{Journal of Statistical Software}, \textbf{23}, 1--46.

\bibitem[{Szeg(1939)}]{szeg1939orthogonal}
Szeg, G. (1939) \textit{Orthogonal polynomials}, vol.~23.
\newblock American Mathematical Soc.

\bibitem[{Willmot(1987)}]{willmot87}
Willmot, G.~E. (1987) The {Poisson-inverse Gaussian} distribution as an
  alternative to the negative {Binomial}.
\newblock \textit{Scandinavian Actuarial Journal}, \textbf{1987}, 113--127.

\bibitem[{Wood(2006)}]{wood2006generalized}
Wood, S. (2006) \textit{Generalized additive models: an introduction with R}.
\newblock CRC press.

\end{thebibliography}

\end{document}